%% file: main.tex
\documentclass[11pt]{article}

\usepackage[final]{acl}

\usepackage{times}
\usepackage{latexsym}
\usepackage{enumitem}
\usepackage{subcaption}
\usepackage{adjustbox}
\usepackage{cuted}
\usepackage{capt-of}
\usepackage{placeins}
\usepackage{float}
\usepackage[T1]{fontenc}

\usepackage[utf8]{inputenc}
\usepackage{dblfloatfix}
\usepackage{microtype}

\usepackage{inconsolata}

\usepackage{graphicx}
\usepackage{cuted}
\usepackage{array}
\usepackage{booktabs}
\usepackage{subcaption}
\usepackage{balance}

\usepackage{xcolor}

\usepackage{tabularx}
\usepackage{xurl}
\usepackage{afterpage}
\usepackage{xspace}
\newif\ifshowcomments
\showcommentstrue

\makeatletter
\let\todo\@undefined
\makeatother

\ifshowcomments
    \usepackage[textsize=scriptsize,textwidth=2.4cm]{todonotes}
\else
    \usepackage[disable,textsize=scriptsize,textwidth=2.4cm]{todonotes}
\fi

\newcommand{\sysname}{OLLA}

\title{Are We There Yet? Assessing Computer-Use Agents for Blind Users’ Accessible Interaction with Desktop Applications}

\author{
\textbf{Satwik Ram Kodandaram\textsuperscript{1}}\thanks{{Correspondence:}
\href{mailto:skodandaram@cs.stonybrook.edu}
     {skodandaram@cs.stonybrook.edu}},
\textbf{Monalika Padma Reddy\textsuperscript{1}},
\textbf{Xiaojun Bi\textsuperscript{1}}, \\
\textbf{Jiawei Zhou\textsuperscript{1}},
\textbf{I. V. Ramakrishnan\textsuperscript{1}},
\textbf{Vikas Ashok\textsuperscript{2}} \\
\\
\textsuperscript{1}Stony Brook University \quad
\textsuperscript{2}Old Dominion University\\
}

\begin{document}

\maketitle

\input{Sections/0.Abstract}

\input{Sections/1.Introduction}
\input{Sections/2.RelatedWork}
\input{Sections/3.Study}
\input{Sections/4.KeyFindings}
\input{Sections/5.Discussion}
\input{Sections/6.Conclusion}
\input{Sections/7.Limitations}
\input{Sections/8.EthicalConsiderations}
\input{Sections/Acknowledgement}

\balance
\bibliography{bibliography}

\input{Sections/Appendix}




\end{document}

%% file: Sections/0.Abstract.tex
\begin{abstract}
Computer-use agents are emerging as a paradigm for agentic human-AI interaction, combining language reasoning with multimodal interface grounding to operate GUIs. Yet their effectiveness for blind screen-reader users in real-world desktop workflows remains unclear. We present a three-week diary study with \(8\) blind users using \sysname{}, a screen-reader-accessible CUA prototype, collecting \(1{,}258\) commands across \(12\) applications with screenshots, UI trees, model responses, and action traces. We evaluate GPT-5 during deployment and re-execute the same commands with four additional models. GPT-5 achieved the highest success rate at \(52.5\%\). Trace analysis reveals grounding, planning, constraint-tracking, and termination failures, while interviews reveal beyond-automation needs.
\end{abstract}


%% file: Sections/1.Introduction.tex
\section{Introduction}

\begin{figure*}[t]
    \centering
    \includegraphics[width=0.65\textwidth]{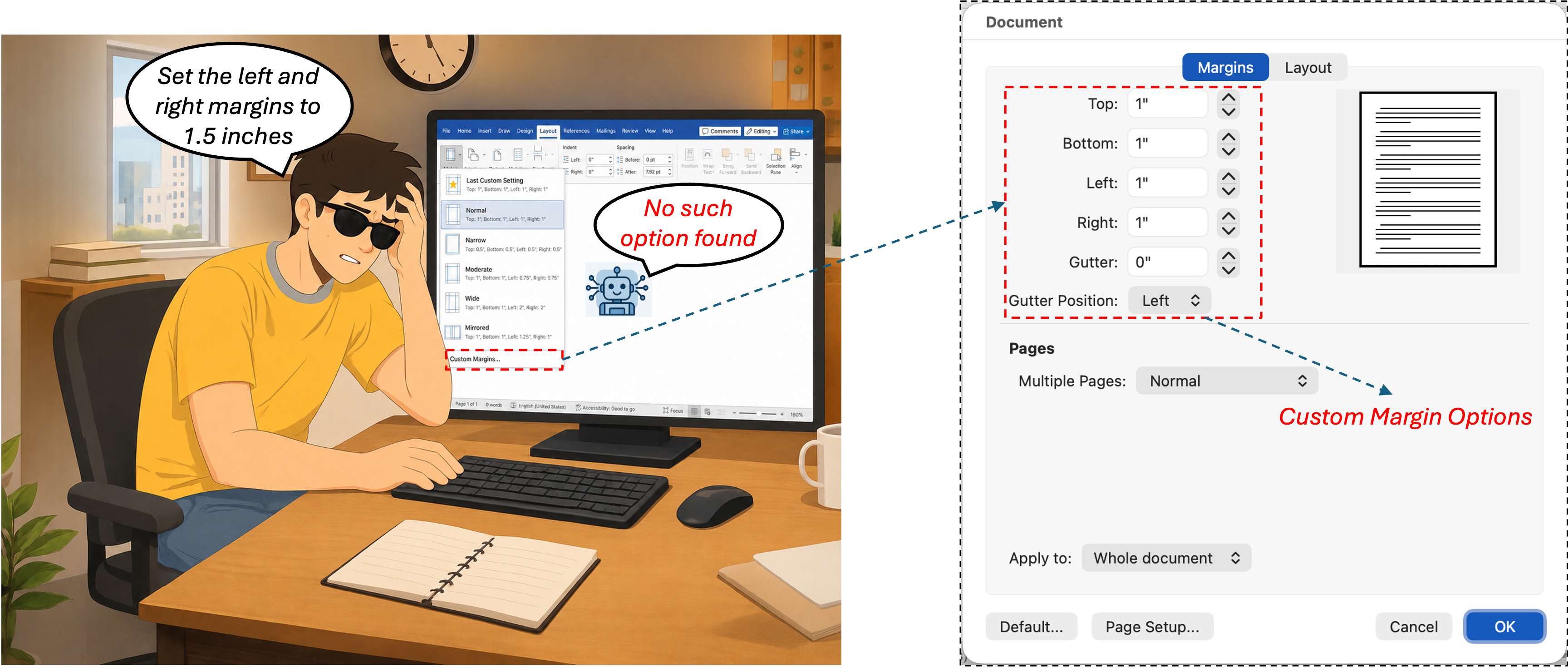}
    \caption{Illustration of a CUA failure in Word. During a custom margin task, the agent failed to find the available margin settings and incorrectly reported that no matching option existed.}
    \label{fig:teaser}
\end{figure*}

Computer-use agents (CUAs), e.g., OpenAI’s Operator~\cite{openai2025operator}, Anthropic’s Computer Use Tool~\cite{anthropic2025computeruse}, Microsoft Copilot~\cite{microsoft2026copilot}, Google DeepMind’s Project Astra~\cite{google2025projectastra}, and Google’s Gemini Spark~\cite{coimbra2026geminispark}, are emerging as multimodal agents that perceive graphical interfaces, reason over user instructions, and execute actions across applications. These systems combine language reasoning with multimodal grounding over screenshots~\cite{zhang2025ufo,zhou2023webarena} and DOM representations~\cite{xie2024osworld} to perform tasks such as navigation, clicking, and typing. Recent benchmarks report rapid progress on web and desktop automation tasks~\cite{davydova2025osuniverse}, positioning CUAs as a promising paradigm for general-purpose human-AI interaction.


CUAs may be particularly valuable for blind users, who often face substantial barriers when interacting with modern graphical user interfaces (GUIs). Blind users typically rely on screen readers such as NVDA~\cite{nvaccess}, JAWS~\cite{jaws}, and VoiceOver~\cite{voiceover}, which vocalize interface content or present it through refreshable braille displays. Interaction is largely sequential and keyboard-driven, often conflicting with GUIs designed around spatial layouts and point-and-click interaction. Prior work shows that this mismatch creates persistent challenges in locating controls, navigating nested structures, understanding dynamic updates, recovering from errors, and adapting to interface changes~\cite{wentz2011usability,ashok2018non,leporini2012interacting,uckun2022taming}.

Recent accessibility research has begun exploring CUAs for nonvisual computer interaction~\cite{gubbi2026a11y,peng2025morae}. However, prior studies have largely relied on simulated evaluations, persona prompting, or narrowly scoped laboratory tasks, leaving a limited understanding of how CUAs perform in real-world nonvisual workflows~\cite{gubbi2026a11y,zhou2026position}. Consequently, it remains unclear how reliably current CUAs support blind users in everyday computer use, what failures emerge in practice, and how blind users envision leveraging these systems beyond automation.

To address this gap, we conducted an IRB-approved three-week longitudinal diary study with $8$ blind screen-reader users interacting with a custom screen-reader-accessible CUA prototype, named \sysname{}. The study investigated the following research questions:

\begin{itemize}[noitemsep,leftmargin=*]
    \item \textbf{RQ1.} How effectively do computer-use agents support blind users in completing everyday, real-world computer tasks?
    \item \textbf{RQ2.} Where do CUAs break down during nonvisual task execution, and what do these breakdowns reveal about current agent limitations?
    \item \textbf{RQ3.} How do blind users envision CUAs supporting everyday computer use beyond end-to-end automation?
\end{itemize}

Because existing CUAs are predominantly visually mediated, we develop \sysname{},\footnote{\url{https://github.com/Satwikram/OLLA}} a screen-reader-friendly interface that enabled participants to issue commands, monitor execution progress, and review agent actions non-visually. \sysname{} functions as an accessibility layer over existing CUAs, enabling blind screen-reader users to interact with them without modifying the underlying agent architecture or reasoning process. During deployment, \sysname{} used GPT-5 to ensure a consistent participant experience. Participants used the system during authentic desktop workflows spanning \(12\) applications, generating \(1{,}258\) blind user-issued commands. For each command, the system recorded (with participants' permission) execution traces including screenshots, UI trees, model responses, generated actions, and interaction histories. We later re-executed the same participant-issued commands using Claude Sonnet, Gemini 2.5 CU, UI-TARS, and Qwen3-VL in fresh live application instances for controlled cross-model analysis.

Our findings show that current CUAs complete a meaningful subset of nonvisual computer tasks but remain unreliable. GPT-5 achieved the highest success rate (\(52.5\%\)), followed by Claude Sonnet (\(48.5\%\)), Gemini 2.5 CU (\(43.9\%\)), UI-TARS (\(39.8\%\)), and Qwen3-VL (\(37.9\%\)). Trace analysis revealed recurring failures in UI grounding \citep{lan2026seeing}, prior-knowledge reliance \citep{li2025okbench,feng2026tracking}, multi-step planning \citep{feng2025unraveling}, constraint tracking \citep{zhou2022online}, and termination behavior \citep{yang2026capability}. Figure~\ref{fig:teaser} shows one such case, where the agent incorrectly concluded that Word lacked the requested custom-margin option despite it being available. Beyond automation performance, interviews show that participants envision CUAs as collaborative support systems for recovering from unfamiliar states, understanding interfaces, troubleshooting errors, learning applications, and improving efficiency in repetitive or technical workflows \citep{kodandaram2026finding}. In summary, this paper makes the following contributions:


\begin{itemize}[noitemsep,leftmargin=*]
    \item We introduce a human-centered dataset of \(1{,}258\) blind user-issued desktop commands collected during a three-week longitudinal deployment, paired with detailed CUA execution traces including screenshots, UI trees, model responses, generated actions, and interaction histories.

    \item We provide a human-centered empirical evaluation of contemporary computer-use agents grounded in blind users' real-world nonvisual desktop workflows, identifying systematic failures in grounding, planning, constraint tracking, and interaction management across multiple large language models.
    
    
    \item We characterize how blind users envision CUAs beyond end-to-end automation, highlighting opportunities for adaptive guidance, interface learning, troubleshooting support, and productivity assistance in accessible computing.
\end{itemize}

%% file: Sections/2.RelatedWork.tex
\section{Background}

\subsection{Evaluating Computer-Use Agents}

Recent work has developed numerous benchmarks for evaluating computer-use agents across web, desktop, operating system, and mobile environments. Web benchmarks such as WebShop, Mind2Web, WebArena, VisualWebArena, WebVoyager, WorkArena, and BrowserGym evaluate language-guided interaction, action prediction, visual grounding, and end-to-end task completion~\cite{yao2022webshop,deng2023mind2web,zhou2023webarena,koh2024visualwebarena,he2024webvoyager,drouin2024workarena,chezelles2024browsergym}. OSWorld and Windows Agent Arena extend evaluation to desktop and operating-system tasks~\cite{xie2024osworld,bonatti2024windows}, while Android in the Wild and AndroidWorld evaluate mobile device-control agents~\cite{rawles2023androidinthewild,rawles2405androidworld}.

Beyond task completion, newer benchmarks examine online realism, workplace autonomy, safety, and accessibility. Online-Mind2Web examines whether offline benchmarks overestimate progress in live web settings~\cite{xue2025illusion}; TheAgentCompany evaluates workplace agents~\cite{xu2026theagentcompany}; ST-WebAgentBench studies safety and policy compliance~\cite{levy2024st}; and BLIND-ACT examines infeasible, ambiguous, or inappropriate goals~\cite{shayegani2025just}. Accessibility-focused work also shows that CUA performance drops under assistive-technology interaction conditions~\cite{gubbi2026a11y}.

Together, these benchmarks show strong progress in evaluating CUA capabilities across web, desktop, mobile, workplace, safety, and accessibility settings. However, they do not fully capture how effectively CUAs support blind users in everyday real-world computer tasks, where they fail, or how future systems should be designed as effective assistive agents. This paper addresses this gap.

\subsection{AI-Mediated Nonvisual Computer Use}

Blind users typically interact with computer applications using screen readers such as NVDA~\cite{nvaccess}, JAWS~\cite{jaws}, and VoiceOver~\cite{voiceover}. Prior work has studied accessibility barriers in web and desktop applications~\cite{doush2013non,islam2023probabilistic,sunkara2023assessing,kodandaram2023detecting} and proposed guidelines for assistive-technology compatibility~\cite{wcag22,waiaria12,harper2012web,morales2013design}. However, accessibility does not necessarily imply usability. Even when controls are technically accessible, blind users may still struggle to locate, understand, and operate them~\cite{wentz2011usability,ashok2018non,leporini2012interacting,uckun2022taming}. These difficulties reflect the mismatch between visually organized GUIs and sequential screen-reader interaction~\cite{wentz2013survey,miao2016contrasting,baldwin2017tangible}, and are further amplified by application heterogeneity, complex shortcuts, and shifting interaction patterns~\cite{billah2017ubiquitous,kodandaram2024enabling}.

Prior systems have reduced these burdens through interface adaptation, structured navigation, context-aware guidance, and uniform interaction mechanisms~\cite{lee2020repurposing,uckun2022taming,chen2026struggle}. LLM-based systems further support natural-language commands and interface automation~\cite{kodandaram2024enabling}, while CUA-focused work highlights accessibility gaps and the need for mixed-initiative interaction~\cite{gubbi2026a11y,peng2025morae}. However, we still lack a clear understanding of real-world effectiveness of CUAs experienced by blind users in everyday computer-use contexts, where CUAs are used and where they fail, and how blind users envision support from CUAs beyond automation.

%% file: Sections/3.Study.tex
\section{Evaluating CUAs as Assistive Agents for Blind Users}


Existing CUA benchmarks, such as OSWorld~\cite{xie2024osworld}, Windows Agent Arena~\cite{bonatti2024windows}, and WebArena~\cite{zhou2023webarena}, evaluate general agent capabilities using well-formed prompts, controlled initial states, and measurable end conditions. However, evaluating CUAs for blind users requires data grounded in everyday nonvisual computer use. Existing benchmarks do not capture how blind users formulate commands or provide enough step-by-step evidence to analyze how failures unfold. To this end, we collect human-centered interaction data from blind users' everyday desktop use, as described next.

\subsection{Human-Centered Data Collection}


To collect data grounded in everyday nonvisual computer use, we conducted an IRB-approved three-week diary study with \(8\) blind screen-reader users. Diary studies capture repeated experiences in naturalistic settings over time~\cite{bolger2003diary,caruana2015longitudinal}, making them appropriate for studying technology use in everyday human-subjects contexts~\cite{lazar2017research}. Participants used our desktop CUA prototype to issue natural-language commands to an agent that could observe, reason about, and act on computer interfaces, allowing us to capture commands, task contexts, breakdowns, and reflections close to the moment of use.

Existing CUAs and open-source GUI agents often require visual monitoring of screenshots, interface changes, or agent actions~\cite{gubbi2026a11y}, making them difficult to deploy directly with blind participants. We therefore developed \sysname{} as a screen-reader-accessible interaction layer over existing CUAs, enabling blind users to issue commands, monitor execution, and review agent actions nonvisually without altering the underlying agent architecture or reasoning process. Participants installed \sysname{} with a \texttt{.exe} installer and used it during regular desktop activities. For each task attempt, \sysname{} logged the participant's command, UI tree, screenshot, model response, generated action, and interaction history. Additional study and system details are in Appendix~\ref{app:study-details}.


\subsection{Cross-Model Replay Evaluation}
\label{sec:replay-evaluation}
To compare CUA performance under controlled conditions, we re-executed participant-issued commands collected during the diary study with each of four additional CUA models. For each model-command pair, we reset the task to its initial state by opening a fresh live application instance and executed the command independently. Models did not interact with reconstructed traces or continue from another model's state, preventing cross-execution side effects. Across models, the \sysname{} pipeline, participant-issued command, system prompt, structured action schema, execution environment, and action executor were held constant. Screenshots, UI trees, and interaction histories were extracted from the application state after each action using the same mechanism for every model.

\subsection{Outcome Annotation and Reference-Step Construction}

Four human annotators annotated the execution data using a shared protocol. They first independently performed each task in the corresponding application state to determine the minimum sequence of task-relevant actions required for successful completion, which was used to construct the reference steps for each command. The annotators then independently evaluated each model's execution trace and labeled the outcome as success, partial completion, or failure, while annotating all applicable failure modes. Inter-annotator agreement was measured using Krippendorff's $\alpha$, yielding $\alpha=0.84$. Disagreements were reviewed collectively and resolved through consensus, and agreed-upon labels were used in the final analysis.

\subsection{Interaction Log and Qualitative Analysis}

We analyzed \sysname{} logs at the task and step levels by reviewing the user command, model response, selected action, UI-tree state, screenshot, and interaction history. This allowed us to examine how the agent interpreted each task, what actions it selected, and where execution succeeded or broke down. We analyzed post-study interview data using hybrid reflexive thematic analysis~\cite{bingham2021deductive,braun2021thematic,naeem2023step}, combining deductive codes guided by our research questions with inductive codes that emerged from participants' responses.

%% file: Sections/4.KeyFindings.tex

\input{Sections/Findings/RQ1}
\input{Sections/Findings/RQ2}
\input{Sections/Findings/RQ3}

%% file: Sections/Findings/RQ1.tex
\section{RQ1. Effectiveness of CUAs}

\begin{figure*}[t]
\centering
\makebox[\textwidth][c]{%
\begin{subfigure}[t]{0.48\textwidth}
    \centering
    \includegraphics[width=\linewidth]{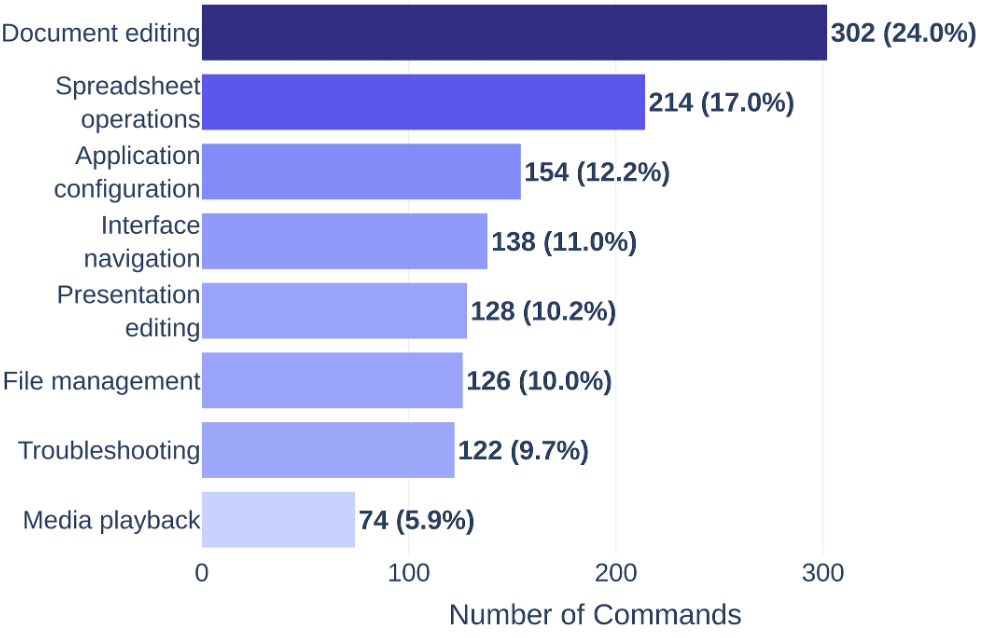}
    \caption{Commands by task category}
    \label{fig:task-category-distribution}
\end{subfigure}
\hspace{0.015\textwidth}
\begin{subfigure}[t]{0.48\textwidth}
    \centering
    \includegraphics[width=\linewidth]{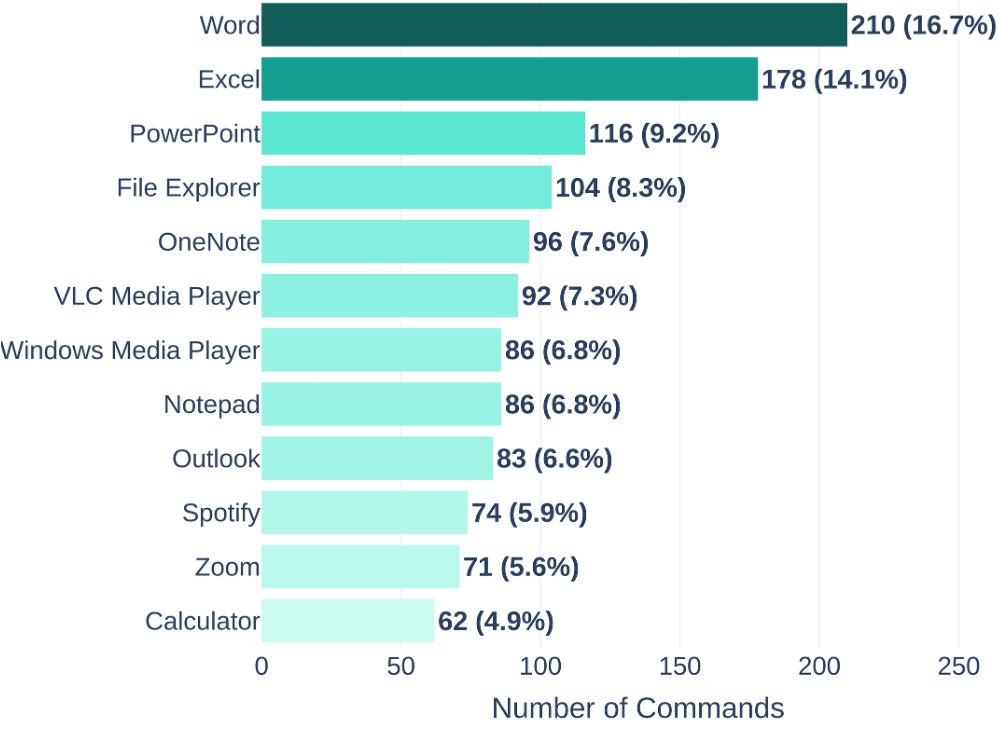}
    \caption{Commands by application}
    \label{fig:application-command-distribution}
\end{subfigure}
}
\caption{Distribution of participant-issued commands across task categories and applications.}
\label{fig:command-distribution}
\end{figure*}



Our evaluation combines in-the-wild use with controlled cross-model re-execution of the same participant-issued commands. We first analyze the deployed study agent during participants' everyday desktop workflows, then compare five contemporary CUAs using the same commands participants issued during the study.

\subsection{Study Agent and Baselines}

During the study, \sysname{} was deployed with GPT-5~\cite{singh2025openai} as the underlying agent to maintain a consistent configuration across participants. We later re-ran the same participant-issued commands through the \sysname{} pipeline using four additional models: Claude Sonnet with Computer Use~\cite{anthropic2025computeruse}, Gemini 2.5 Computer Use~\cite{google2025computeruse}, UI-TARS~\cite{qin2025ui}, and Qwen3-VL~\cite{bai2025qwen3}.\footnote{Exact model identifiers and execution configurations for all evaluated models are reported in Appendix~\ref{app:model-configurations}, Table~\ref{tab:model-configurations}.} These models cover complementary CUA directions, including proprietary computer-use agents, specialized GUI agents, and open multimodal models with visual-agent capabilities.

Although some models were designed primarily for screenshot-based perception, we evaluate all models with the same task command, UI tree, screenshot, and recent interaction history. The UI tree provides semantic information about controls, roles, and hierarchy, complementing screenshots, which alone can be challenging for precise GUI grounding and coordinate prediction~\cite{lin2025showui,qin2025ui}. Each model generates the same structured user-interface action output, which is executed through Microsoft UI Automation~\cite{microsoft_uiautomation_win32}.

\subsection{Participant-Issued Commands}

Across the three-week study, participants issued \(N=1{,}258\) commands to \sysname{} across \(12\) desktop applications. We manually grouped collected commands within each application that represented the same underlying task despite differences in wording or parameter values, resulting in \(304\) normalized task intents. For example, ``\textit{insert a table with 4 by 4 cells}'' and ``\textit{insert a table with 15 by 15 cells}'' share the same intent, inserting a table. Using inductive qualitative content analysis~\cite{hsieh2005three}, we grouped the \(304\) normalized intents into eight broader task categories derived from the collected commands. Figure~\ref{fig:task-category-distribution} shows the category distribution, and Figure~\ref{fig:application-command-distribution} shows the application distribution, with the largest numbers from Word, Excel, PowerPoint, and OneNote.


\subsection{Task Outcomes and Step Progress}

We treat each participant-issued command and execution trace as one task attempt, coded as \textit{success}, \textit{partial completion}, or \textit{failure}. Success means full task completion; partial completion means completing at least one required task-relevant step without finishing the task; and failure means no valid task progress, an incorrect outcome, or a repeated non-progressing loop.

Figure~\ref{fig:model-performance} and Table~\ref{tab:model-performance} in Appendix~\ref{app:performance-tables} show that GPT-5 has the highest observed success rate at \(52.5\%\) (95\% CI: \(49.8\)--\(55.3\%\)), followed by Claude Sonnet at \(48.5\%\) (95\% CI: \(45.7\)--\(51.3\%\)), Gemini 2.5 CU at \(43.9\%\) (95\% CI: \(41.2\)--\(46.6\%\)), UI-TARS at \(39.8\%\) (95\% CI: \(37.2\)--\(42.6\%\)), and Qwen3-VL at \(37.9\%\) (95\% CI: \(35.3\)--\(40.6\%\)). Although GPT-5 has the highest observed success rate, its \(4.0\)-percentage-point advantage over the next-best model, Claude Sonnet, is not statistically significant in a paired McNemar test (\(\chi^2=3.25\), \(p=.071\)).

Partial completion was common across models, ranging from \(33.3\%\) to \(34.9\%\). We further computed \(\textit{StepProgress}_i=c_i/r_i\), where \(r_i\) is the number of required reference steps and \(c_i\) is the number completed before failure. As shown in Table~\ref{tab:step-progress-partial}, partial traces often involved substantial progress, with GPT-5 completing \(68.3\%\) of required steps on average before breakdown and Qwen3-VL completing \(53.0\%\). Application-level \(\textit{StepProgress}\) results are provided in Table~\ref{tab:app-step-progress}.



\begin{table}[t]
\centering
\small
\setlength{\tabcolsep}{4pt}

\begin{tabular}{@{}lcc@{}}
\toprule
\textbf{Model} & \textbf{Partial Completion Traces} & \textbf{Step Progress} \\
\midrule
GPT-5 & 431 & 68.3\% \\
Claude & 429 & 65.6\% \\
Gemini & 439 & 60.6\% \\
UI-TARS & 430 & 56.2\% \\
Qwen3-VL & 419 & 53.0\% \\
\bottomrule
\end{tabular}
\begin{minipage}{0.96\linewidth}
\footnotesize
\textit{Note.} Step Progress is the average fraction of reference steps completed before breakdown.
\end{minipage}
\caption{Step-level progress among partially completed commands before breakdown.}
\label{tab:step-progress-partial}
\end{table}
\vspace{-3.25pt}

\begin{figure*}[t]
\centering
\begin{subfigure}[t]{0.43\textwidth}
    \centering
    \includegraphics[width=\linewidth]{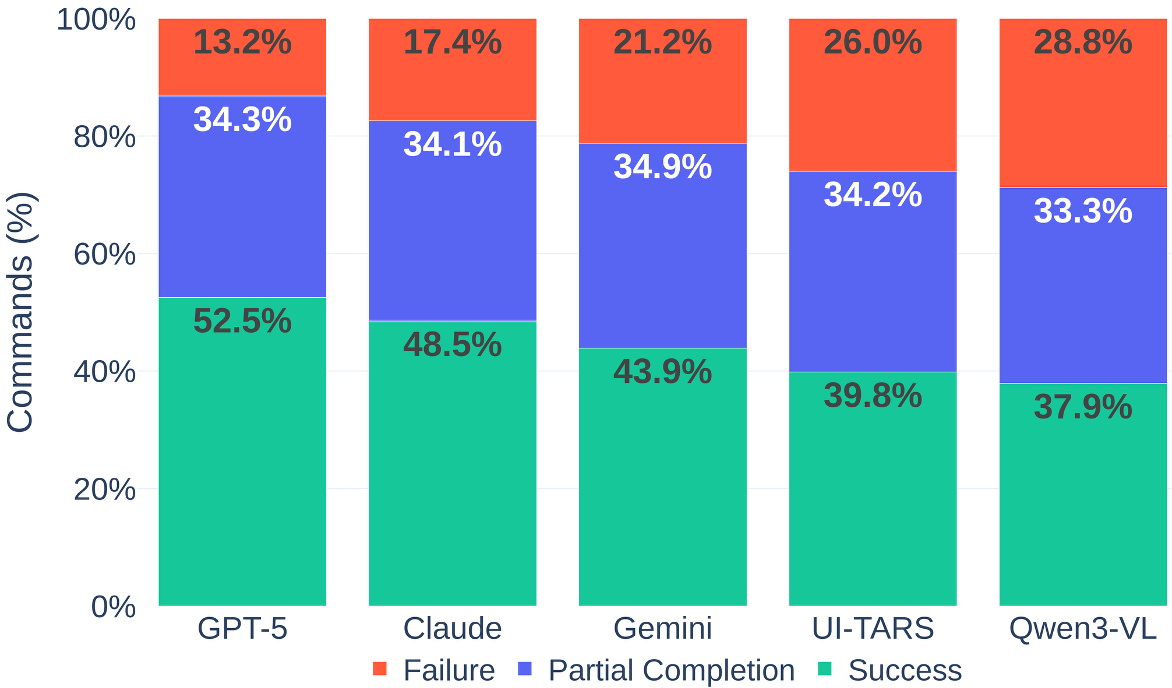}
    \caption{Task-level outcomes (Table~\ref{tab:model-performance} in Appendix~\ref{app:performance-tables})}
    \label{fig:overall-model-outcomes}
\end{subfigure}
\hfill
\begin{subfigure}[t]{0.55\textwidth}
    \centering
    \includegraphics[width=\linewidth]{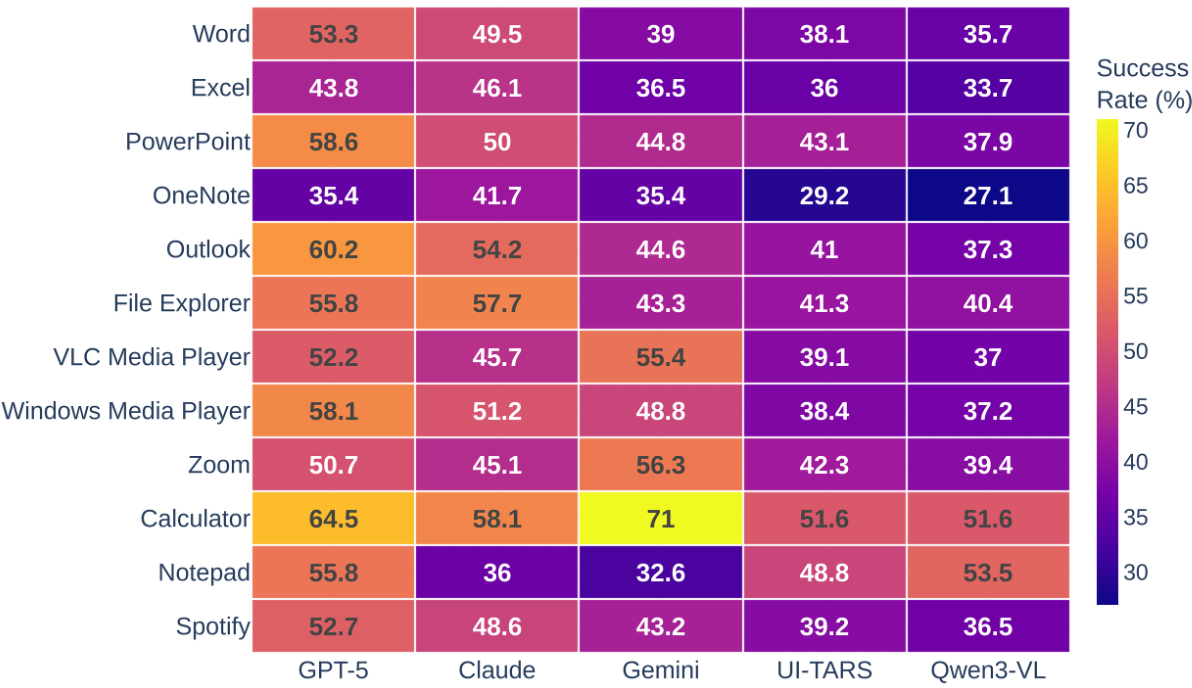}
    \caption{Application-level success rate (Table~\ref{tab:app-level-performance} in Appendix~\ref{app:performance-tables})}
    \label{fig:application-success-heatmap}
\end{subfigure}
\caption{CUA performance across models.}
\label{fig:model-performance}
\end{figure*}

\subsection{Interactions CUAs Handled More Successfully}

Successful interactions were concentrated in tasks with direct mappings between user commands and readily identifiable interface controls. These included direct property modifications (e.g., changing font family, font size, or text formatting), single-step interface operations (e.g., inserting tables, comments, or page breaks), navigation and information-retrieval tasks (e.g., opening menus, switching tabs, or locating settings), and simple content-editing tasks with explicitly specified parameters. In contrast, tasks requiring multi-step reasoning, discovery of hidden controls, or maintenance of multiple constraints were less consistently successful. These patterns characterize capabilities demonstrated under the observed conditions rather than reliable performance across all task instances or interface states.

\subsection{Application-Level Variation}

We also analyzed outcomes by application. Figure~\ref{fig:model-performance}b shows application-level success rates, while the full success, partial-completion, and failure counts for each model and application are provided in Table~\ref{tab:app-level-performance} (Appendix~\ref{app:performance-tables}). Application-level results show that performance varied across application contexts, suggesting that CUA effectiveness depends not only on the model but also on application structure, control visibility, and task type. These results motivate our trace-based analysis in RQ2, where we examine why agents failed to convert partial progress into full task completion.

The observed success rates characterize only tasks participants chose to attempt. Post-study interviews indicated that some participants avoided sensitive tasks (e.g., banking, passwords, or personal documents) and occasionally stopped attempting task types after repeated failures with similar interactions. For example, failures with advanced formatting in Microsoft Word led some participants to avoid comparable editing tasks in OneNote. Accordingly, the reported success rates should be interpreted as conditional on the tasks participants chose to attempt and do not capture tasks they considered but elected not to delegate to the agent.

%% file: Sections/Findings/RQ2.tex
\begin{figure*}[t]
\centering

\begin{subfigure}[c]{0.47\textwidth}
    \centering
    \adjustbox{valign=c}{%
        \includegraphics[
            width= \linewidth,
            height = 4.5cm,
        ]{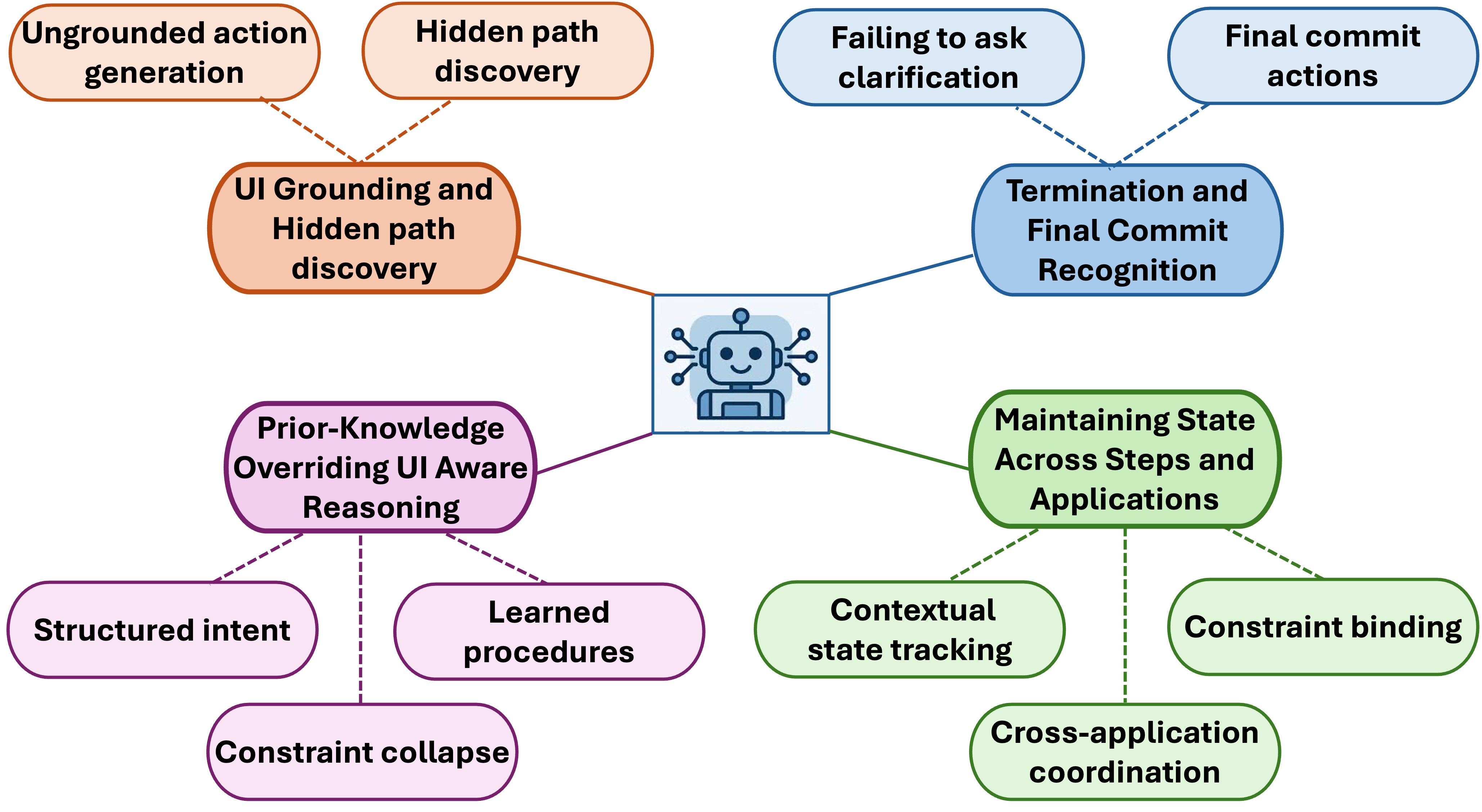}
    }
    \caption{Key failure modes observed in CUAs during nonvisual desktop interaction}
    \label{fig:failure-taxonomy}
\end{subfigure}
\hfill
\begin{subfigure}[c]{0.52\textwidth}
    \centering
    \adjustbox{valign=c}{%
        \includegraphics[
            width=\linewidth,
        ]{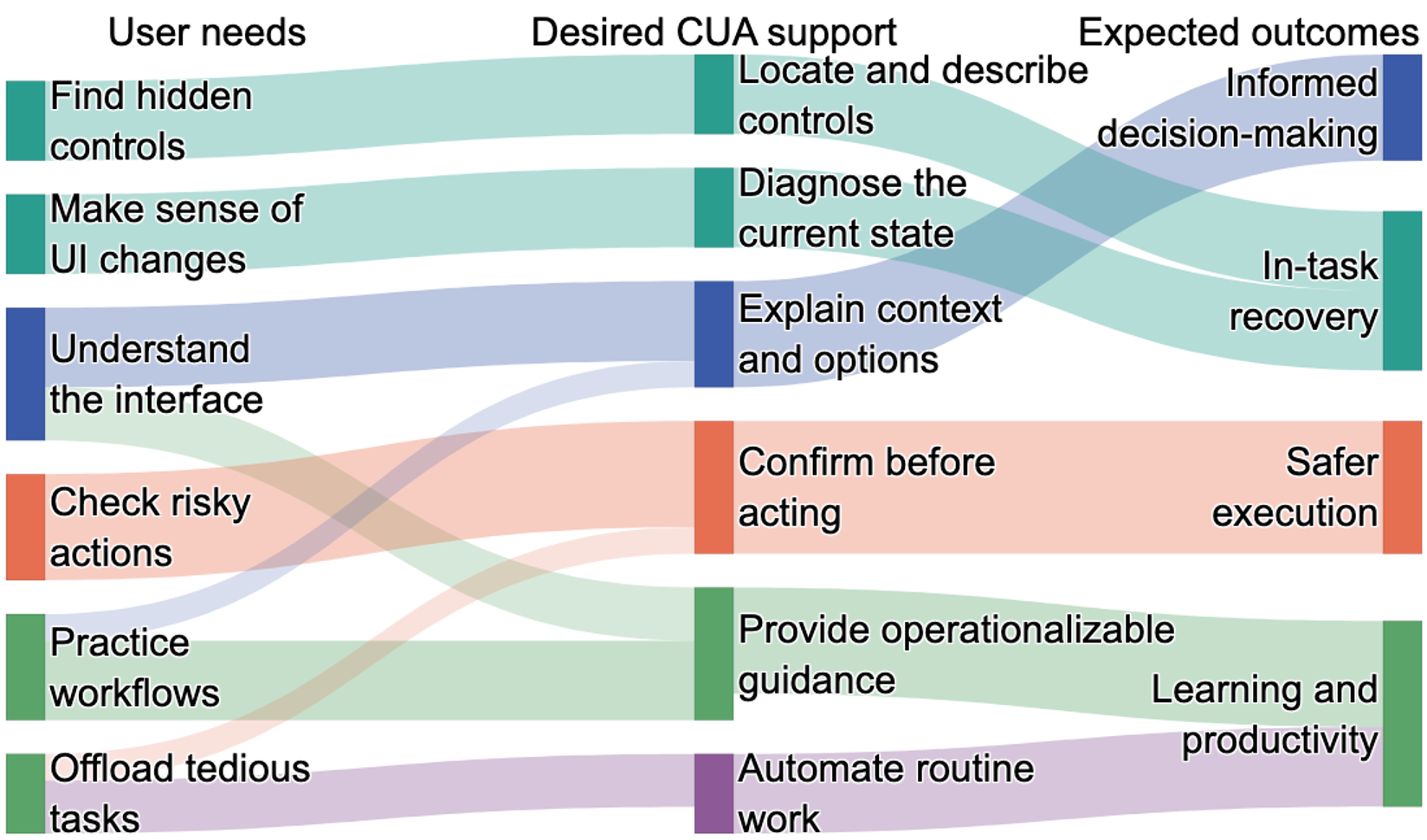}
    }
    \caption{Participant-envisioned support pathways for CUAs}
    \label{fig:sankey}
\end{subfigure}
\caption{Taxonomy of CUA breakdowns and participant-envisioned support pathways.}
\label{fig:break-downs-envision}
\end{figure*}

\section{RQ2. Breakdowns During Nonvisual Task Execution}

As summarized in Figure~\ref{fig:failure-taxonomy}, we group recurring breakdowns into categories including UI grounding, prior-knowledge reliance, multi-step planning, constraint tracking, and termination behavior. For each category, we report how often it appears among the relevant unsuccessful traces.

\subsection{UI Grounding and Hidden Path Discovery}

\paragraph{\textit{Ungrounded action generation}.}
Grounding errors were a common source of non-completion. Among unsuccessful traces, they accounted for \(24.6\%\) of GPT-5 failures (\(147/597\), study only) and \(22.6\%\) of failures across all models (\(789/3{,}489\), study + replays). In these cases, agents generated actions that were not supported by the current UI state, such as hallucinated control names, fabricated control types, or coordinates that did not align with the intended element. These errors were frequent in ribbon-based applications such as \textit{Word}, \textit{Excel}, and \textit{OneNote}, where relevant controls were not always visible. For example, for ``\textit{Protect this document with a password},'' the model placed \textit{Protect Document} under the \textit{Home} tab, although the correct path required the \textit{Review} tab.

\paragraph{\textit{Hidden path discovery}.}
A related failure involved tasks whose target controls were reachable only through intermediate navigation. Among unsuccessful traces, hidden-path failures accounted for \(21.4\%\) of GPT-5 failures (\(128/597\)) and \(20.7\%\) of failures across all models (\(722/3{,}489\)). Agents often handled directly visible options but failed on deeper variants of the same task. For example, they could complete ``\textit{change margin to narrow}'' in \textit{Word}, where \textit{Narrow} appears after opening the \textit{Margins} menu, but failed on custom-margin requests that required opening a dialog and filling multiple fields. Similarly, for ``\textit{insert a table 15 by 15},'' agents stopped at the visible grid limit rather than opening the custom table option. These results suggest that CUAs can act on exposed controls but struggle when task completion depends on discovering hidden interface paths.

\subsection{Prior Knowledge Overriding UI-Aware Reasoning}

\paragraph{\textit{Learned procedure and default-driven errors}.}

Some failures occurred when agents followed familiar procedures or defaults rather than reasoning from the observed UI and command. Among unsuccessful attempts, learned procedure reliance accounted for \(17.1\%\) of GPT-5 failures (\(102/597\)) and \(9.6\%\) across all models (\(335/3{,}489\)). Commands succeeding in \textit{Word} sometimes failed in \textit{OneNote} when the model applied Word-like procedures. For ``\textit{Insert a comment}'' in \textit{Word}, the model followed Microsoft's documented \textit{Review} tab procedure,\footnote{\url{https://shorturl.at/1kkBS}} although the option was available under \textit{Home}. A related pattern was default collapse, where agents selected common options despite constraints. For custom margins, agents sometimes selected \textit{Normal} or \textit{Narrow} rather than opening the custom dialog. This pattern accounted for \(11.4\%\) of GPT-5 failures (\(68/597\)) and \(6.5\%\) across models (\(227/3{,}489\)).

\paragraph{\textit{Structured intent failures}.}
Structured intent failures were less frequent but important for tasks requiring exact formulas, ranges, operators, grouping, or ordering. They accounted for \(5.9\%\) of GPT-5 failures (\(35/597\)) and \(5.7\%\) across all models (\(199/3{,}489\)). For example, ``\textit{In D2, calculate the average of B2 through B10}'' required preserving the target cell, function, and range, but agents sometimes generated an incomplete or incorrect formula. These failures reflected errors in forming the intermediate representation before execution, rather than locating controls. This reinforces prior work showing that blind users often need effortful verification when using generative AI for structured content such as spreadsheets~\cite{perera2026m}.

\subsection{State Maintenance Across Task Steps}
Beyond finding the right controls, agents also needed to preserve task state across steps. Several failures occurred after initially correct actions, when agents lost user constraints or prior context.

\paragraph{\textit{Constraint binding}.}
Constraint-binding failures occurred after agents made partial progress but lost one or more user-specified requirements. Among partial-completion task attempts, they accounted for \(20.6\%\) of GPT-5 cases (\(89/431\)) and \(22.4\%\) across all models (\(481/2{,}148\)). For example, for ``\textit{change the font to Arial and the font size to 14},'' agents sometimes changed the font family but dropped the size constraint. For ``\textit{insert a footer with the page number centered},'' they inserted the page number but failed to preserve centered alignment. These cases explain why partial completions could show substantial step progress while still failing to satisfy the full command.

\paragraph{\textit{Contextual state tracking}.}
Other failures involved losing track of prior interface context. Among unsuccessful task attempts, contextual state tracking accounted for \(10.2\%\) of GPT-5 failures (\(61/597\)) and \(8.7\%\) across all models (\(304/3{,}489\)). In Excel, for ``\textit{create a new sheet and switch back to the previous tab},'' the agent created the new sheet but failed to identify which sheet had been active before creation. Similar failures occurred after pop-ups, subwindows, or transient modes, where the agent needed to resume an earlier working context.

\paragraph{\textit{Cross-application coordination}.}
Commands spanning multiple applications introduced another state-maintenance challenge. Cross-application coordination accounted for \(7.4\%\) of GPT-5 failures (\(44/597\)) and \(6.0\%\) across all models (\(209/3{,}489\)). For ``\textit{copy the chart from Excel and paste it into Word},'' agents often completed only one side of the task, such as copying content in Excel but failing to switch to Word or pasting in the wrong location. These failures show that CUAs need persistent task-state representations that track remaining constraints, prior context, active objects, and source and destination applications across steps.

\subsection{Termination and Commit Recognition}

\paragraph{\textit{Termination recognition}.}
In \sysname{}, task completion was signaled when the agent generated a \textit{done} output. Termination failures occurred when agents stopped too early, continued without meaningful state change, or repeated the same action instead of recognizing that execution was no longer progressing. Among partial-completion attempts, this pattern accounted for \(17.6\%\) of GPT-5 cases (\(76/431\)) and \(19.1\%\) across all models (\(410/2{,}148\)). In one Excel trace, the agent created a new sheet but failed to emit \textit{done}, repeatedly activating the new-sheet control and creating extra sheets. These cases help explain why some attempts showed high step progress but still failed to satisfy the command.

\paragraph{\textit{Final commit actions}.}
Other late-stage failures occurred when agents reached the correct interface path but missed the final action needed to apply the change. For example, in \textit{Word}, for ``\textit{Save this file as a PDF},'' the agent navigated to the export option and selected \textit{PDF}, but failed to click the final \textit{Save} button that completed the export. Here, the failure was not path discovery, but recognizing the commit point where the selected option takes effect. These failures suggest that CUAs need stronger mechanisms for detecting terminal states, non-progressing loops, and confirmation actions.


%% file: Sections/Findings/RQ3.tex
\section{RQ3. CUAs Beyond Automation}

RQ3 examines how participants envisioned CUAs beyond full automation. Figure~\ref{fig:sankey} summarizes these desired roles, showing CUAs as tools for understanding interfaces, getting situated help, controlling risky actions, and learning workflows during nonvisual computer use.

\subsection{Understanding Before Acting}

Participants envisioned CUAs as useful before execution, especially for understanding task context and deciding what to do next. They wanted agents to explain visual content, document structure, and interface elements, such as charts, layouts, controls, menus, dialogs, and settings. As P6\footnote{P1--P8 denote participant identifiers; participant demographics are reported in Table~\ref{tab:study-participants}.} explained, ``\textit{I may not want it to do the whole task for me.}'' This was particularly relevant for unfamiliar or visually organized interfaces, where participants wanted to understand not only what was present, but how it related to the task. Rather than immediately delegating, users wanted CUAs to describe what is relevant to the goal and available options. This could help them decide whether to act manually, request guidance, or ask the agent to execute.

\subsection{Situated Help and Troubleshooting}

Participants also wanted CUAs to help when they were already in the middle of a task and became stuck. In these moments, they did not necessarily want the agent to take over the full task. Instead, they wanted targeted support for locating hidden controls, understanding dialogs or menus, identifying what changed after an action, troubleshooting unexpected states, and deciding the next step. P3 described this need for situated assistance: ``\textit{Sometimes I just need it to tell me where I am, what options are available, and what I should do next.}'' This suggests that CUAs should support opportunistic assistance during nonvisual workflows, allowing users to request explanations, guidance, and recovery support while remaining engaged in the task.

\subsection{User-Controlled Execution}

Participants wanted CUAs to keep them involved during ambiguous, consequential, or difficult-to-verify tasks. Rather than proceeding silently, agents should explain planned actions, ask for missing information, and confirm assumptions before acting. This was especially important for hard-to-undo actions, such as changing settings, modifying files, deleting content, or handling sensitive information. Participants were cautious about uses involving personal or financial data. As P7 emphasized, ``\textit{I want it to ask me before clicking something important. I do not want it making decisions on its own.}'' These concerns suggest that CUAs should support user-controlled execution through confirmations, explanations, and options to approve, modify, or stop actions before commitment.

\subsection{Learning and Productivity Support}

Participants also envisioned CUAs as tools for learning and productivity. They wanted agents to explain the application structure, provide screen-reader-relevant steps, and help them practice workflows independently. As P5 explained, ``\textit{If it can explain how the application is organized and guide me through the steps, I can learn to do it myself the next time}.'' They also saw value in using CUAs for tedious or technically demanding tasks, such as formatting, file management, screen-reader configuration, add-on installation, and customization workflows. These responses suggest that CUAs can support not only immediate task completion, but also longer-term confidence in nonvisual computer use when they help users understand workflows rather than only execute them.

%% file: Sections/5.Discussion.tex
\section{Discussion and Future Work}

Our findings show the promise and current limits of CUAs for everyday nonvisual desktop support. We next discuss implications for designing CUAs that go beyond automation, support clarification and recovery, and better reflect blind users' needs.

\subsection{Interaction-Rich Training Data for Assistive CUAs}

Many failures occurred when agents continued execution instead of pausing, asking for clarification, or adapting to corrections. Our dataset offers a starting point for studying these behaviors by capturing blind user-issued commands, UI states, screenshots, model responses, generated actions, and interaction histories from nonvisual workflows. Future work can extend it with annotations for uncertainty, clarification opportunities, rejected actions, recovery attempts, and decision changes. Prior work shows the value of supervision~\cite{ross2011reduction,christiano2017deep,ouyang2022training,cui2023no,li2026orce}. Semi-automatic tools could convert these logs into training and evaluation data for clarification, backtracking, and recovery.

\subsection{Backtrack via Reward-Guided Execution}

Our failure analysis suggests opportunities for reward-guided improvement. Agents often chose the wrong path, repeated actions, or continued executing despite little progress. Prior work on reinforcement learning from human feedback, preference-based optimization, and process supervision shows that supervision can target both final outcomes and intermediate judgments about useful, safe, or correct behavior~\cite{christiano2017deep,ziegler2019fine,ouyang2022training,lightman2024let,song2025prmbench}. For CUAs, reward models could combine task completion with signals such as goal progress, relevant-control discovery, repeated-action avoidance, constraint preservation, and recovery. For nonvisual use, rewards should also capture clarification, state explanation, and caution around actions difficult to verify or undo.

\subsection{Assistive CUAs for Learning}
Participants' reflections suggest that CUAs should support beyond end-to-end automation. Many nonvisual workflows require targeted help when users are stuck, troubleshooting, or learning unfamiliar interfaces. Future systems could build on natural-language control, interface adaptation, context-aware guidance, and mixed-initiative support~\cite{kodandaram2024enabling,lee2020repurposing,uckun2022taming,lin2025llm,chen2026struggle,peng2025morae}. CUAs could describe the screen, identify controls, suggest screen-reader-friendly steps, ask clarification questions, and offer recovery options. They could also turn action plans into guidance, helping users practice workflows and remain in control during task execution.

%% file: Sections/6.Conclusion.tex
\section{Conclusion}
This paper evaluate computer-use agents in real-world contexts as assistive systems for blind screen-reader users in everyday desktop workflows. Through a three-week diary study and cross-model evaluation of \(1{,}258\) participant-issued commands, we found that CUAs complete meaningful tasks but remain unreliable for nonvisual use. Trace analysis exposed recurring failures in grounding, hidden-path discovery, state maintenance, constraint preservation, and termination recognition. Interviews showed that participants wanted CUAs not as autonomous replacements, but as collaborative support for understanding interfaces, recovering from breakdowns, and learning workflows.

%% file: Sections/7.Limitations.tex
\clearpage
\section*{Limitations}

Our study did have a few limitations, which we discuss below. These limitations reflect the scope of our participant sample, prototype implementation, and evaluation design, and should be considered when interpreting the findings.

\paragraph{Participant Pool.} Our participant pool was limited to blind people who primarily use screen readers for computer interaction. We did not include low-vision users or people with other visual impairment conditions who may rely on different assistive tools, such as screen magnifiers, high-contrast settings, or combined visual and nonvisual strategies. As a result, our findings primarily reflect screen-reader-mediated desktop use and may not fully generalize to users with different access needs or assistive technology practices. Our study also did not include participants under the age of \(18\), whose computer-use practices, learning needs, and support expectations may differ.

\paragraph{Prototype and Operating System Scope.}

Our prototype was implemented for Windows using Microsoft UI Automation tool, which provided structured UI-tree information such as control labels, roles, coordinates, and hierarchy. As a result, our findings reflect CUA behavior in Windows desktop environments and may not fully generalize to macOS or other operating systems, where accessibility APIs, permission models, screen-reader behavior, and the completeness of exposed interface structure can differ~\cite{uckun2022taming,kodandaram2024enabling}. Some platforms may not expose an equivalent UI tree, or may expose less complete or less consistently available interface metadata to automation clients. Future work should examine CUA performance across platforms with different accessibility infrastructures.

\paragraph{Language Scope.}

Our study was conducted in English, including participant commands, system feedback, surveys, and interviews. Although modern LLM-based agents can process multiple languages, non-English and mixed-language use may introduce different command formulations, localization issues, interface-label mismatches, and screen-reader interaction patterns. For example, users may combine English application labels with commands in another language, or use localized versions of applications where menu names and shortcut conventions differ. These factors could shape both agent performance and user expectations in ways not captured by our study. Future work should examine CUAs for nonvisual computer use across languages, localized applications, and multilingual screen-reader workflows.

\paragraph{Application Scope.} 

Our evaluation focused on desktop applications and did not include web applications. Web environments introduce different interaction challenges because page structure is exposed through the DOM or accessibility tree, which may not always reflect the complete or current interface state. Many websites rely on JavaScript, AJAX, infinite scrolling, and interaction-triggered loading, where content is rendered only after scrolling, expanding menus, submitting forms, or activating controls. As a result, some relevant content or controls may be hidden behind scripts and may not appear in the available structure until specific interactions occur. Although screenshots can provide complementary visual context, screenshot-only or OCR-based representations can introduce recognition errors that affect grounding and action generation. Future work should examine CUAs for nonvisual use across dynamic web applications, where DOM structure, accessibility-tree information, and visual state may diverge.

\paragraph{Model Scope.}

Our evaluation was limited to five agent-capable models. We selected these models to cover a range of current CUA-relevant capabilities, including frontier proprietary models, a computer-use-specialized model, and open-weight multimodal baselines that could be integrated into our evaluation pipeline. We did not evaluate every available model because each model run required executing \(1{,}258\) commands with step-by-step UI observations, generated actions, and logged traces, making the evaluation costly in terms of compute, API usage, infrastructure, and manual outcome verification. Because CUAs and multimodal models are rapidly evolving, future models may show different strengths or failure patterns. Our goal was therefore not to provide a definitive ranking of models, but to characterize current capabilities and recurring breakdowns in nonvisual desktop use.

%% file: Sections/8.EthicalConsiderations.tex
\section*{Ethical Considerations}

This study was approved by our institutional review board (IRB). Because participants were blind screen-reader users, we designed recruitment, consent form, installation, study instructions, surveys, and interviews to be accessible with screen readers. Participants were informed about the study purpose, duration, data collection procedures, and their ability to stop participation or skip any task. We framed all outcomes as evaluations of CUA behavior rather than user performance, since failures could reflect agent limitations, interface accessibility issues, or both.

The study involved privacy risks because CUA traces can include screenshots, UI trees, user commands, model outputs, and interaction history from participants' personal computers. We therefore treated logs as potentially sensitive data. We minimized unnecessary collection, anonymized participant identifiers, and used secure storage with access limited to the research team. In reporting findings, we describe task patterns and failures without revealing personally identifying content.

%% file: Sections/Acknowledgement.tex
\section*{Acknowledgement}
This work was supported by NIH Award R01EY035688 and DoD Award HT94252410098. Jiawei Zhou is supported by an Amazon Research Award on AWS Agentic AI and a Stony Brook OVPR Seed Grant.


%% file: Sections/Appendix.tex
\clearpage
\appendix

\input{Sections/Appendix/Study}

\input{Sections/Appendix/EvaluationDetails}

\input{Sections/Appendix/Tables}

\input{Sections/Appendix/FailureExamples}

%% file: Sections/Appendix/Study.tex
\section{Study and System Details}
\label{app:study-details}

\subsection{Participant Recruitment and Eligibility}

We recruited participants from an existing contact list maintained from prior IRB-approved accessibility studies. We contacted individuals who had previously consented to be recontacted and had indicated interest in research on accessibility and technology use. We also used snowball sampling~\cite{noy2008sampling}, inviting participants to share the study with other eligible peers. In accordance with our IRB protocol, we conducted outreach using each participant’s preferred communication method, such as email or phone.

\begin{table*}[!t]
\centering
\footnotesize
\setlength{\tabcolsep}{5pt}
\renewcommand{\arraystretch}{1.35}

\begin{tabular}{@{}
>{\centering\arraybackslash}m{0.8cm}
>{\centering\arraybackslash}m{1.4cm}
>{\centering\arraybackslash}m{1.6cm}
>{\centering\arraybackslash}m{1.5cm}
>{\centering\arraybackslash}m{1.6cm}
>{\centering\arraybackslash}m{5.0cm}
>{\centering\arraybackslash}m{2.15cm}
@{}}
\toprule
\textbf{ID} & \textbf{Age / Gender} & \textbf{Vision Loss (Onset / LP)} & \textbf{Preferred Screen Reader} & \textbf{Expertise} & \textbf{Familiar Applications} & \textbf{Computer Usage Frequency} \\
\midrule
P1  & 59 / Male   & 28 / No  & JAWS & Beginner     & MS Word, Google Chrome, Zoom & Bi-Weekly \\ \hline
P2  & 34 / Female & 19 / Yes & NVDA & Intermediate & MS Word, MS Excel, MS PowerPoint, MS Teams, Mozilla Firefox & Daily \\ \hline
P3  & 30 / Male   & 18 / Yes & JAWS & Expert       & MS Word, MS Excel, Zoom, PyCharm, MS PowerPoint, Google Chrome & Weekly \\ \hline
P4  & 24 / Female & 2 / No   & NVDA & Intermediate & MS Word, Zoom, Mozilla Firefox & Daily \\ \hline
P5  & 59 / Male   & 39 / No  & JAWS & Beginner     & MS Word, MS Excel, Mozilla Firefox, Zoom, Notepad++, MS PowerPoint & Weekly \\ \hline
P6  & 44 / Female & 17 / No  & NVDA & Expert       & MS Word, MS Excel, MS PowerPoint, Google Chrome, MS Teams, VS Code & Daily \\ \hline
P7  & 68 / Female & 19 / Yes & JAWS & Intermediate & MS Word, Zoom, Google Chrome & Daily \\ \hline
P8  & 37 / Female & 3 / Yes  & NVDA & Expert       & MS Word, MS Excel, Zoom, MS PowerPoint, VS Code, Mozilla Firefox & Daily \\
\bottomrule
\end{tabular}

\caption{Summary of participant demographics, onset of visual impairment, preferred screen reader, self-reported expertise, desktop applications used, and frequency of computer use.}
\label{tab:study-participants}
\end{table*}

Participants were eligible if they self-identified as blind, relied on a screen reader for regular computer use, and had prior experience using desktop applications (e.g., Word, Excel) for everyday tasks. Because the study focused on computer-use agents for desktop interaction, participants also needed access to a personal computer and sufficient familiarity with common applications to use the study system during the three-week diary period. Participants needed to be able to communicate in English. We excluded individuals under \(18\), those who did not regularly use a screen reader, and those who lacked prior experience with everyday computer tasks. We confirmed eligibility through a brief screening interview.

Overall, \(N=8\) blind screen-reader users completed the study. During the three-week study period, participants used the CUA-based application for everyday computer tasks across \(12\) different desktop applications. All participants reported regular screen reader use and prior experience using desktop applications for everyday tasks. Participant demographics are summarized in Table~\ref{tab:study-participants}. Participants received \$100 in compensation for their time and contributions.

\subsection{Procedure}

The study spanned three weeks. Participants were given an executable (\texttt{.exe}) installer for \sysname{} and instructions for installing and launching the application on their own computers. To holistically evaluate \sysname{} across diverse interaction scenarios, participants were encouraged to use the system as part of their everyday computer activities and to issue at least $10$ commands per day across desktop applications, including tasks such as configuring settings, managing files, navigating interfaces, and troubleshooting issues.

Participants used \sysname{} in their own environments. After each task attempt, participants had the option to complete a brief survey describing the task, the outcome, and any additional comments, including major issues or breakdowns they encountered. We encouraged participants to report both successful and unsuccessful experiences.

We conducted periodic check-ins during the study to address technical issues and ensure continued participation without disruption. At the end of the study, we conducted semi-structured interviews to gather additional insights into participants’ experiences, including how they used \sysname{}, where it succeeded or failed, and how it fit into their everyday workflows.

All participants provided informed consent prior to participation. Participants were informed about the types of interaction data collected during system use, including screenshots, UI trees, model responses, action traces, and survey/interview responses. Participants could stop participation at any time, and all collected data were anonymized with personally identifiable information removed prior to analysis.

\subsection{Post-Study Interviews}

After the three-week diary period, we conducted semi-structured interviews with each participant to gather deeper reflections on their experience using \sysname{}. The interviews were guided by participants' interaction logs and post-task survey responses, which allowed us to discuss specific task attempts where \sysname{} succeeded, partially completed the task, or failed. This helped us interpret log-coded breakdowns by asking what made the agent's behavior helpful, confusing, incomplete, or difficult to monitor nonvisually.

The interviews also directly informed our analysis of how blind users envisioned CUAs beyond end-to-end automation. We asked participants how they imagined using CUAs in everyday computing activities, when they would want guidance, clarification, troubleshooting support, learning support, or productivity assistance, and what expectations they had around reliability, user control, privacy, and trust.

\subsection{\sysname{} Design and Implementation}

We developed \sysname{} because existing CUAs and open-source GUI agents are difficult to deploy directly with blind participants. Many systems are designed around visually mediated interaction, where users monitor screenshots, pointer movements, or visual state changes, and they often provide limited screen-reader access, limited keyboard-based control, or limited access to step-by-step execution logs~\cite{gubbi2026a11y}.

\sysname{} was developed as a screen-reader-friendly CUA prototype for collecting interaction data from blind users. Its design was informed by recent CUAs and GUI agents that follow an observation-action loop, where the agent observes the interface, reasons over the user goal, predicts an action, executes it, and observes the updated state~\cite{openai2025operator,anthropic2025computeruse,zhang2025ufo,xie2024osworld,bonatti2024windows}. It was also informed by accessibility-focused LLM systems for nonvisual computer support, which use UI-tree representations to capture accessibility properties such as control labels, roles, hierarchy, focusable elements, states, and element coordinates~\cite{kodandaram2024enabling,chen2026struggle,gubbi2026a11y}. Following these systems, we crafted \sysname{}'s prompts to combine the user's natural-language command with the current interface state, recent interaction history, and a structured output format. \sysname{} uses both screenshots and the Microsoft UI Automation tree at each execution step. Screenshots provide rendered visual context, while the UI tree grounds this context in semantically exposed interface structure. This combination reduces reliance on screenshots alone, which can miss or misread interface content due to OCR errors, visually similar controls, missing labels, or ambiguous layouts.

Once launched, \sysname{} runs in the background and can be activated through a global keyboard shortcut, allowing users to invoke the agent without visually locating the application window. Users issue natural-language commands, after which \sysname{} follows an iterative perceive-reason-act loop. At each step, \sysname{} extracts the current screenshot and UI tree, combines them with the user command and recent interaction history, and sends this context to the LLM. The model outputs a brief decision rationale and a structured JSON action specification, including the action type, target control label, control role, element coordinates \((x,y,w,h)\), and text content when needed. When the model determines that the task is complete, it returns a \texttt{done} output. \sysname{} then executes the predicted action through Microsoft UI Automation, including clicking, typing, scrolling, or selecting controls~\cite{microsoft_uiautomation_win32}.

To support nonvisual monitoring, \sysname{} provides audio feedback after each action and after determining that the task is complete. During each task attempt, \sysname{} stores the user command, model rationale, structured action output, executed action, screenshot, UI-tree state, and interaction history. We used LangSmith~\cite{langsmith2026} to log these traces and maintain a bounded interaction buffer, allowing the agent to condition later actions on recent context. These logs supported our later analysis of task outcomes and breakdowns. Finally, all collected data was anonymized.

%% file: Sections/Appendix/EvaluationDetails.tex
\section{Evaluation and Data Details}
\label{app:evaluation-details}

\subsection{Dataset Composition and Task Distribution}
\label{app:command-distribution}

\begin{table*}[t]
\centering
\small
\begin{tabular}{p{0.17\textwidth} p{0.43\textwidth} r r}
\toprule
\textbf{Category} & \textbf{Definition} & \textbf{Commands} & \textbf{\%} \\
\midrule
Document Editing &
Creating, modifying, formatting, or organizing content in text-based documents. &
302 & 24.0 \\

Spreadsheet Operations &
Creating, editing, formatting, or computing over spreadsheet data, cells, formulas, or charts. &
214 & 17.0 \\

Application Configuration &
Changing application preferences, settings, or configuration options. &
154 & 12.2 \\

Interface Navigation &
Locating or navigating among controls, menus, tabs, dialogs, views, or other interface elements. &
138 & 11.0 \\

Presentation Editing &
Creating or modifying slides, slide content, layouts, or presentation formatting. &
128 & 10.2 \\

File Management &
Creating, locating, opening, saving, moving, renaming, exporting, or deleting files and folders. &
126 & 10.0 \\

Troubleshooting &
Diagnosing or resolving application, configuration, or interaction problems. &
122 & 9.7 \\

Media Playback &
Controlling or configuring audio or video playback. &
74 & 5.9 \\
\midrule
\textbf{Total} & & \textbf{1,258} & \textbf{100.0} \\
\bottomrule
\end{tabular}
\caption{Distribution and operational definitions of the eight task categories derived from participant-issued commands.}
\label{tab:task-category-details}
\end{table*}

Across the study, participants issued \(1{,}258\) commands, which were consolidated into \(304\) normalized task intents. Thus, \(954\) commands represented additional instances of these intents through repeated requests, alternative phrasings, or different parameter values, corresponding to an average of \(4.14\) commands per normalized intent. Participant-level command counts were approximately: P1 (\(158\)), P2 (\(149\)), P3 (\(171\)), P4 (\(136\)), P5 (\(162\)), P6 (\(155\)), P7 (\(148\)), and P8 (\(179\)).

\begin{table*}[t]
\centering
\small
\setlength{\tabcolsep}{5pt}
\begin{tabularx}{\textwidth}{
    >{\raggedright\arraybackslash}p{0.14\textwidth}
    >{\raggedright\arraybackslash}p{0.27\textwidth}
    >{\raggedright\arraybackslash}X}
\toprule
\textbf{Model} & \textbf{Model Identifier} & \textbf{Configuration} \\
\midrule

GPT-5 &
\url{gpt-5-2025-08-07} &
OpenAI API; multimodal screenshot and text input; Microsoft UI Automation tree and interaction history supplied in the prompt; structured JSON actions following the shared \sysname{} action schema. \\

Claude Sonnet &
\url{claude-sonnet-4-6} &
Anthropic API with computer-use capability; screenshot, UI tree, command, and recent interaction history provided at each step; outputs mapped to the shared \sysname{} action schema. \\

Gemini 2.5 CU &
\url{gemini-2.5-computer-use-preview-10-2025} &
Gemini Computer Use API; multimodal screenshot and text input; model-generated interface actions mapped to the common \sysname{} execution interface. \\

UI-TARS &
\url{ByteDance-Seed/UI-TARS-1.5-7B} &
Open-weight multimodal GUI agent; screenshot and textual task context provided at each step; generated GUI actions translated to the shared \sysname{} action schema. \\

Qwen3-VL &
\url{Qwen/Qwen3-VL-8B-Instruct} &
Open-weight multimodal instruction model; screenshot, task command, UI information, and interaction history provided through the same \sysname{} pipeline; structured actions produced using the common action schema. \\

\bottomrule
\end{tabularx}

\caption{Models and execution configurations used in the cross-model evaluation. All models used a maximum output length of 2,048 tokens.}
\label{tab:model-configurations}
\end{table*}

Using inductive qualitative content analysis~\cite{hsieh2005three}, we grouped the normalized intents into eight task categories derived from the collected commands rather than defined a priori. Table~\ref{tab:task-category-details} summarizes each category and its distribution.

\subsection{Model and Execution Configurations}
\label{app:model-configurations}

Table~\ref{tab:model-configurations} summarizes the models used in our evaluation. Exact model versions were fixed for the evaluation, and all models operated through the common \sysname{} pipeline described in Section~\ref{sec:replay-evaluation}.

Across models, the participant-issued command, system prompt, structured action schema, execution environment, and action executor were held fixed. Model-specific adaptations were limited to formatting inputs and outputs according to each model's API or inference interface.

\subsection{Statistical Analysis}
\label{app:statistical-analysis}

To account for uncertainty in the observed model success rates, we computed 95\% Wilson confidence intervals and conducted paired command-level comparisons. GPT-5 achieved a success rate of \(52.5\%\) (95\% CI: \(49.8\)--\(55.3\%\)), compared with \(48.5\%\) for Claude Sonnet (95\% CI: \(45.7\)--\(51.3\%\)). A paired McNemar test found that this \(4.0\)-percentage-point difference was not statistically significant (\(\chi^2=3.25\), \(p=.071\)). Thus, although GPT-5 had the highest observed success rate, the difference from the next-best model should not be interpreted as evidence of a definitive performance advantage.

\subsection{Data and Research Artifacts}
\label{app:artifacts}

The \sysname{} implementation is publicly available on GitHub.\footnote{\url{https://github.com/Satwikram/OLLA}} The anonymized dataset, including participant-issued commands, model outputs, and turn-by-turn execution traces and annotations, is available in our \href{https://github.com/Satwikram/OLLA-Diary-Study}{data repository}.

%% file: Sections/Appendix/Tables.tex
\FloatBarrier
\clearpage
\onecolumn
\section{Task Outcomes by Model and Application}
\label{app:performance-tables}

\begin{table}[H]
\centering
\small
\begin{tabular}{lcccc}
\toprule
\textbf{Model} & \textbf{Success} & \textbf{95\% CI} &
\textbf{Partial} & \textbf{Failure} \\
\midrule
GPT-5 &
661 (52.5\%) & 49.8--55.3 &
431 (34.3\%) & 166 (13.2\%) \\

Claude Sonnet &
610 (48.5\%) & 45.7--51.3 &
429 (34.1\%) & 219 (17.4\%) \\

Gemini 2.5 CU &
552 (43.9\%) & 41.2--46.6 &
439 (34.9\%) & 267 (21.2\%) \\

UI-TARS &
501 (39.8\%) & 37.2--42.6 &
430 (34.2\%) & 327 (26.0\%) \\

Qwen3-VL &
477 (37.9\%) & 35.3--40.6 &
419 (33.3\%) & 362 (28.8\%) \\
\bottomrule
\end{tabular}

\caption{Task-level performance across models on participant-issued commands (\(N=1{,}258\)). Confidence intervals are 95\% Wilson intervals for success proportions.}
\label{tab:model-performance}
\end{table}

\begin{table}[H]
\centering
\footnotesize
\setlength{\tabcolsep}{3.5pt}
\renewcommand{\arraystretch}{1.15}

\resizebox{\textwidth}{!}{%
\begin{tabular}{l r ccc ccc ccc ccc ccc}
\toprule
\textbf{Application} & \textbf{Cmds.}
& \multicolumn{3}{c}{\textbf{GPT-5}}
& \multicolumn{3}{c}{\textbf{Claude}}
& \multicolumn{3}{c}{\textbf{Gemini}}
& \multicolumn{3}{c}{\textbf{UI-TARS}}
& \multicolumn{3}{c}{\textbf{Qwen3-VL}} \\
\cmidrule(lr){3-5}
\cmidrule(lr){6-8}
\cmidrule(lr){9-11}
\cmidrule(lr){12-14}
\cmidrule(lr){15-17}
& & \textbf{S} & \textbf{PC} & \textbf{F}
& \textbf{S} & \textbf{PC} & \textbf{F}
& \textbf{S} & \textbf{PC} & \textbf{F}
& \textbf{S} & \textbf{PC} & \textbf{F}
& \textbf{S} & \textbf{PC} & \textbf{F} \\
\midrule

Word & 210
& \textbf{112} & 71 & 27
& 104 & 70 & 36
& 82 & 79 & 49
& 80 & 73 & 57
& 75 & 73 & 62 \\

Excel & 178
& 78 & 72 & 28
& \textbf{82} & 64 & 32
& 65 & 70 & 43
& 64 & 65 & 49
& 60 & 63 & 55 \\

PowerPoint & 116
& \textbf{68} & 35 & 13
& 58 & 39 & 19
& 52 & 40 & 24
& 50 & 38 & 28
& 44 & 39 & 33 \\

OneNote & 96
& 34 & 45 & 17
& \textbf{40} & 37 & 19
& 34 & 39 & 23
& 28 & 39 & 29
& 26 & 38 & 32 \\

Outlook & 83
& \textbf{50} & 24 & 9
& 45 & 25 & 13
& 37 & 29 & 17
& 34 & 28 & 21
& 31 & 28 & 24 \\

File Explorer & 104
& 58 & 33 & 13
& \textbf{60} & 29 & 15
& 45 & 37 & 22
& 43 & 35 & 26
& 42 & 33 & 29 \\

VLC Media Player & 92
& 48 & 32 & 12
& 42 & 33 & 17
& \textbf{51} & 26 & 15
& 36 & 32 & 24
& 34 & 31 & 27 \\

Windows Media Player & 86
& \textbf{50} & 26 & 10
& 44 & 28 & 14
& 42 & 27 & 17
& 33 & 30 & 23
& 32 & 29 & 25 \\

Zoom & 71
& 36 & 25 & 10
& 32 & 26 & 13
& \textbf{40} & 19 & 12
& 30 & 23 & 18
& 28 & 23 & 20 \\

Calculator & 62
& 40 & 16 & 6
& 36 & 17 & 9
& \textbf{44} & 11 & 7
& 32 & 17 & 13
& 32 & 16 & 14 \\

Notepad & 86
& \textbf{48} & 27 & 11
& 31 & 37 & 18
& 28 & 36 & 22
& 42 & 25 & 19
& 46 & 21 & 19 \\

Spotify & 74
& \textbf{39} & 25 & 10
& 36 & 24 & 14
& 32 & 26 & 16
& 29 & 25 & 20
& 27 & 25 & 22 \\

\midrule
\textbf{Total} & \textbf{1,258}
& \textbf{661} & \textbf{431} & \textbf{166}
& \textbf{610} & \textbf{429} & \textbf{219}
& \textbf{552} & \textbf{439} & \textbf{267}
& \textbf{501} & \textbf{430} & \textbf{327}
& \textbf{477} & \textbf{419} & \textbf{362} \\

\bottomrule
\end{tabular}%
}
\caption{Application-level outcome breakdown across models. S denotes success, PC denotes partial completion, and F denotes failure. Counts are reported for each application; bold indicates the highest success count for each application.}
\label{tab:app-level-performance}
\end{table}

\begin{table}[H]
\centering
\footnotesize
\setlength{\tabcolsep}{5pt}
\renewcommand{\arraystretch}{1.15}

\resizebox{\textwidth}{!}{%
\begin{tabular}{lccccc}
\toprule
\textbf{Application}
& \textbf{GPT-5}
& \textbf{Claude}
& \textbf{Gemini}
& \textbf{UI-TARS}
& \textbf{Qwen3-VL} \\
\midrule
Word & 69.2\% (71) & 66.1\% (70) & 61.4\% (79) & 56.8\% (73) & 53.5\% (73) \\
Excel & 67.5\% (72) & 65.2\% (64) & 60.1\% (70) & 55.7\% (65) & 52.9\% (63) \\
PowerPoint & 70.1\% (35) & 66.8\% (39) & 61.9\% (40) & 57.2\% (38) & 53.6\% (39) \\
OneNote & 65.8\% (45) & 63.7\% (37) & 58.6\% (39) & 54.4\% (39) & 51.7\% (38) \\
Outlook & 69.7\% (24) & 66.3\% (25) & 61.2\% (29) & 56.5\% (28) & 53.2\% (28) \\
File Explorer & 68.9\% (33) & 65.9\% (29) & 60.7\% (37) & 56.1\% (35) & 52.8\% (33) \\
VLC Media Player & 66.4\% (32) & 63.8\% (33) & 58.9\% (26) & 54.7\% (32) & 51.8\% (31) \\
Windows Media Player & 66.9\% (26) & 64.2\% (28) & 59.3\% (27) & 54.9\% (30) & 52.1\% (29) \\
Zoom & 67.8\% (25) & 64.6\% (26) & 59.7\% (19) & 55.3\% (23) & 52.4\% (23) \\
Calculator & 70.5\% (16) & 67.1\% (17) & 62.4\% (11) & 57.5\% (17) & 54.1\% (16) \\
Notepad & 68.2\% (27) & 64.9\% (37) & 59.6\% (36) & 55.4\% (25) & 52.5\% (21) \\
Spotify & 66.7\% (25) & 63.9\% (24) & 57.4\% (26) & 54.6\% (25) & 51.4\% (25) \\
\midrule
\textbf{Overall} & \textbf{68.3\% (431)} & \textbf{65.6\% (429)} & \textbf{60.6\% (439)} & \textbf{56.2\% (430)} & \textbf{53.0\% (419)} \\
\bottomrule
\end{tabular}%
}

\caption{Application-level step progress among partially completed commands. Each cell reports the average fraction of reference steps completed before breakdown, with the number of partial-completion traces in parentheses.}
\label{tab:app-step-progress}
\end{table}

\newpage

%% file: Sections/Appendix/FailureExamples.tex
\section{Visualizing CUA Failure Modes}
\begin{figure}[H]
    \centering

    \begin{subfigure}[t]{0.48\textwidth}
        \centering
        \includegraphics[width=\linewidth]{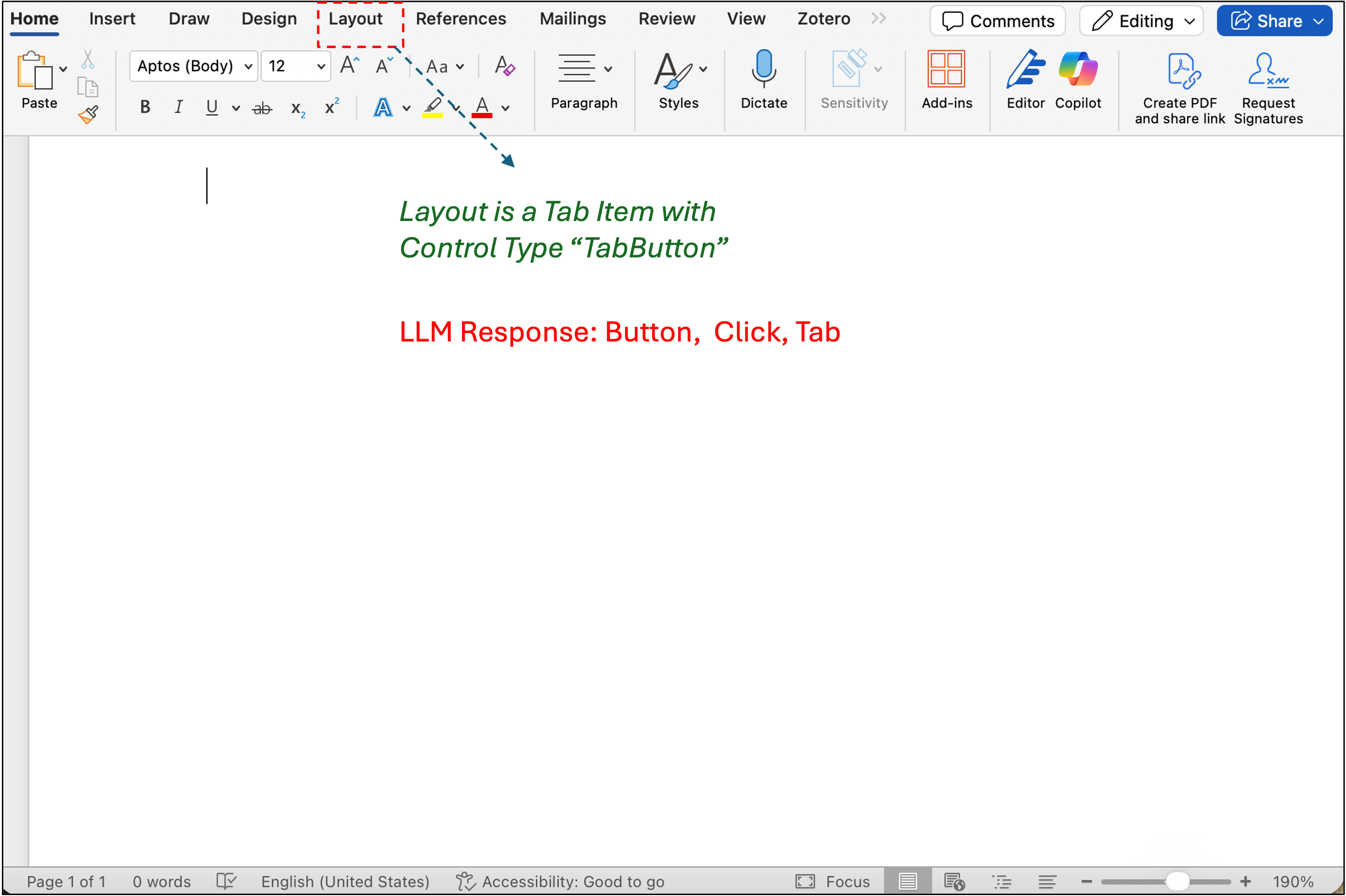}
        \caption{\textbf{Incorrect UI grounding.}
        The UI representation identifies \textit{Layout} as a
        \texttt{TabButton}, whereas the model generates an incompatible
        control representation in its structured action. The resulting
        action is therefore inconsistent with the interface state
        available to the model.}
        \label{fig:failure-grounding}
    \end{subfigure}
    \hfill
    \begin{subfigure}[t]{0.48\textwidth}
        \centering
        \includegraphics[width=\linewidth]{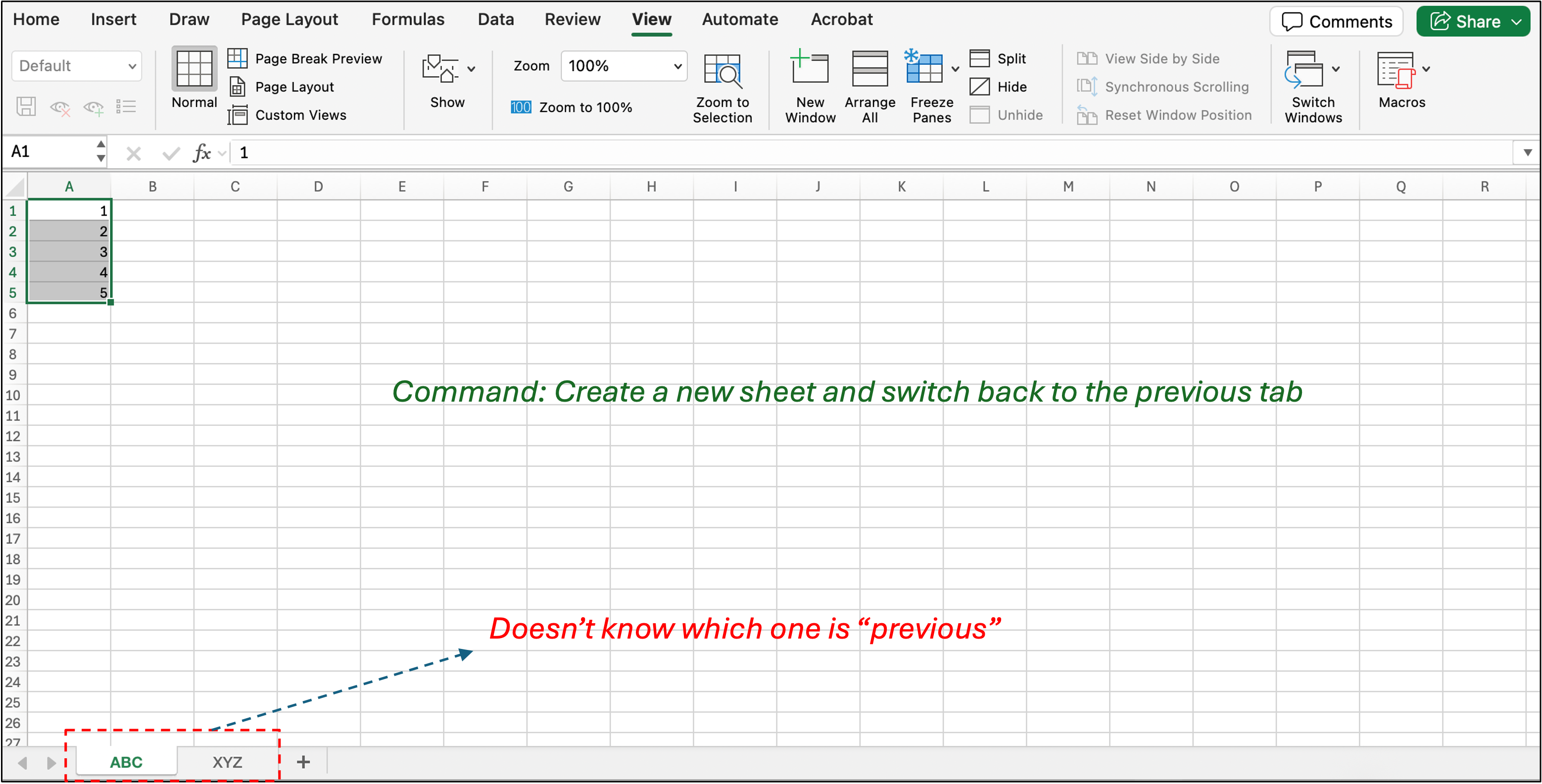}
        \caption{\textbf{Contextual state tracking.}
        After creating a new worksheet, the agent no longer retains
        which sheet was active beforehand and consequently cannot resolve
        the user's reference to the ``previous tab.''}
        \label{fig:failure-context}
    \end{subfigure}

    \caption{\textbf{Breakdowns in interface grounding and state
    maintenance.} The left trace shows an action specification that
    conflicts with the observed UI representation; the right trace shows
    loss of task-relevant interface context across successive actions.}
    \label{fig:failure-examples-grounding}
\end{figure}

\begin{figure}[H]
    \centering

    \begin{subfigure}[t]{0.48\textwidth}
        \centering
        \includegraphics[width=\linewidth]{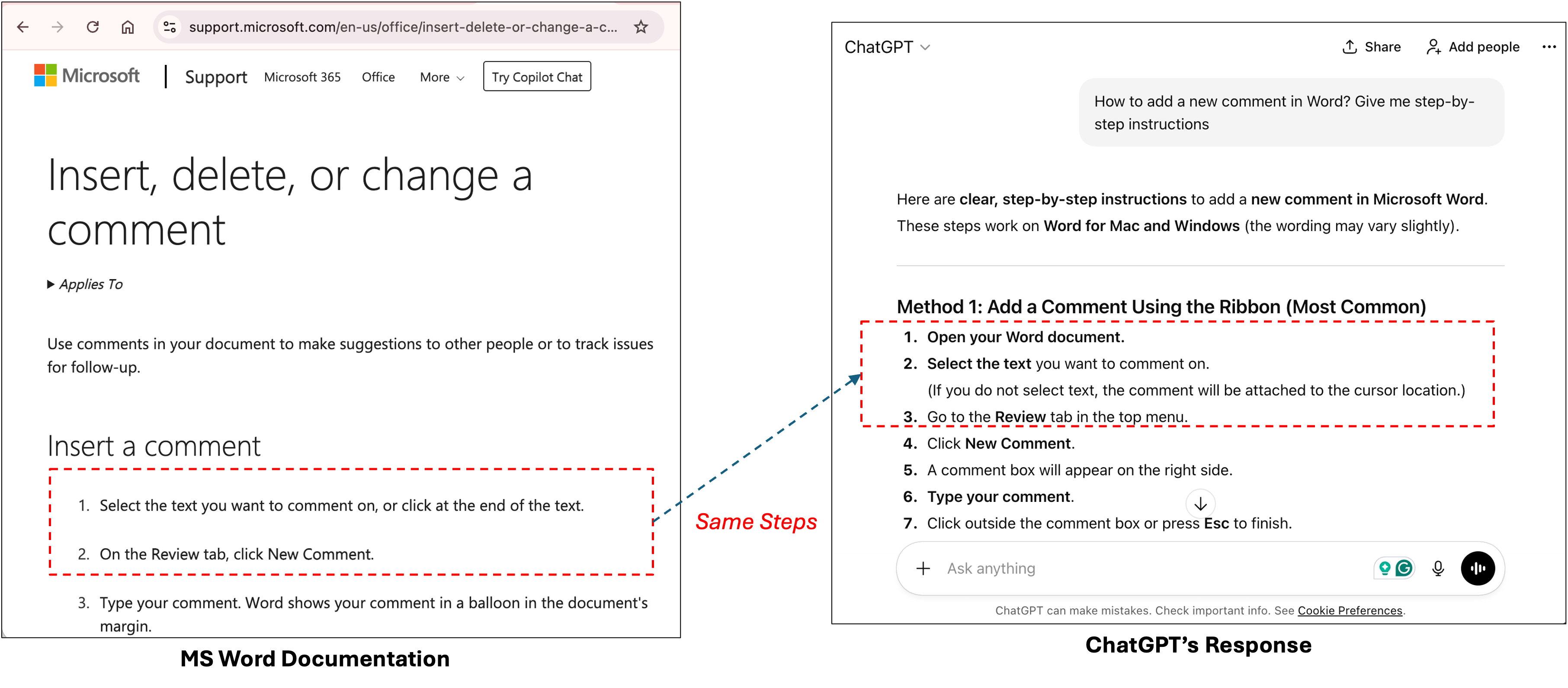}
        \caption{\textbf{Prior-knowledge reliance.}
        The model follows the conventional Microsoft-documented path
        through the \textit{Review} tab for inserting a comment, even
        though the observed interface exposes the relevant control
        directly under \textit{Home}. Learned procedural knowledge
        overrides reasoning from the current UI state.}
        \label{fig:failure-prior}
    \end{subfigure}
    \hfill
    \begin{subfigure}[t]{0.48\textwidth}
        \centering
        \includegraphics[width=\linewidth]{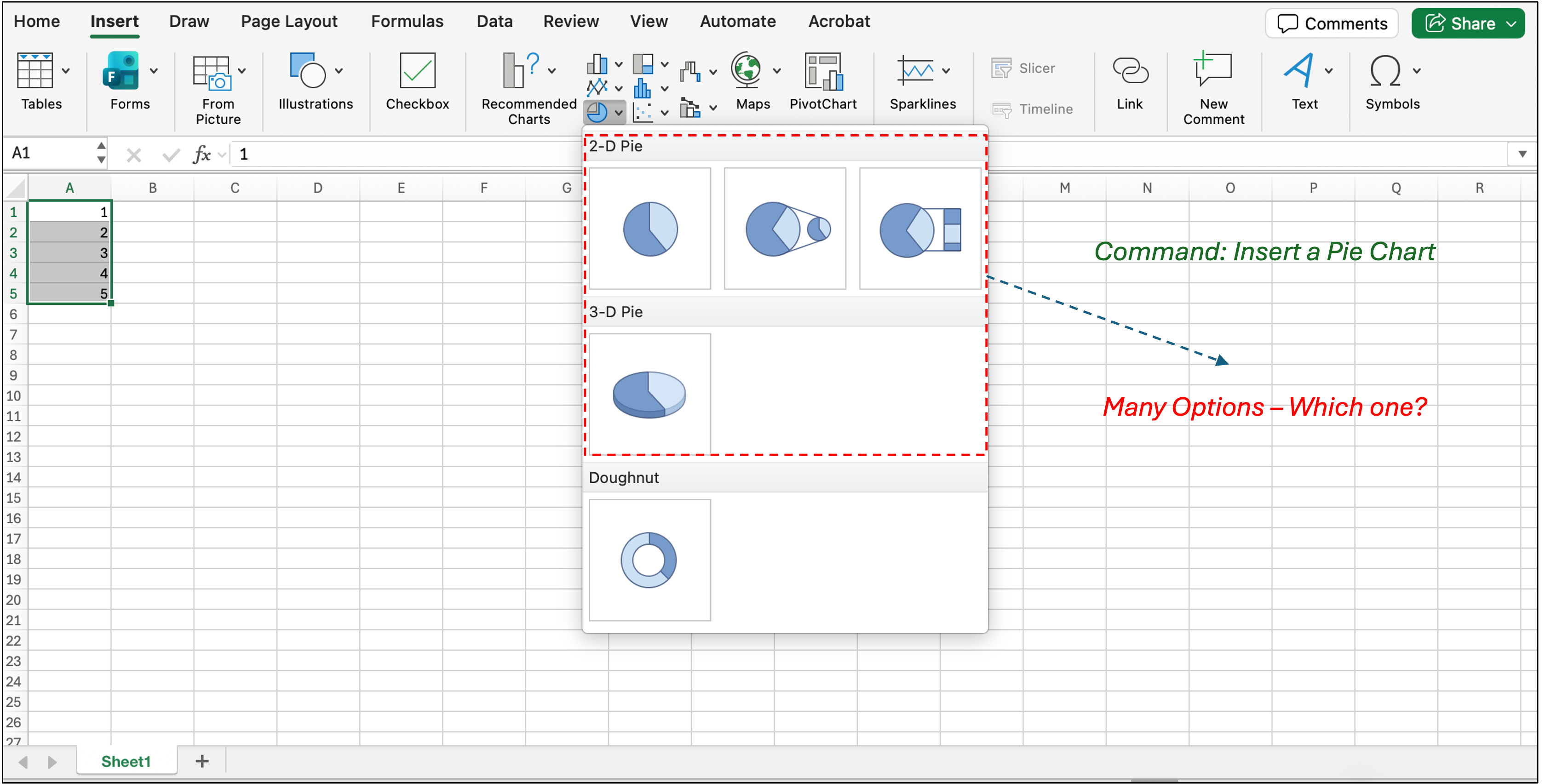}
        \caption{\textbf{Default action selection.}
        The user requests a \textit{3D} pie chart and the interface
        exposes both 2D and 3D variants, yet the agent selects the first
        default pie-chart option rather than preserving the specified
        chart type.}
        \label{fig:failure-default}
    \end{subfigure}

    \caption{\textbf{Breakdowns in situated reasoning and intent
    preservation.} The left trace illustrates reliance on a familiar
    application procedure despite contradictory interface evidence,
    whereas the right trace shows an explicit user constraint being
    collapsed into a readily available default.}
    \label{fig:failure-examples-reasoning}
\end{figure}

\begin{figure}[H]
    \centering

    \begin{subfigure}[t]{0.48\textwidth}
        \centering
        \includegraphics[width=\linewidth]{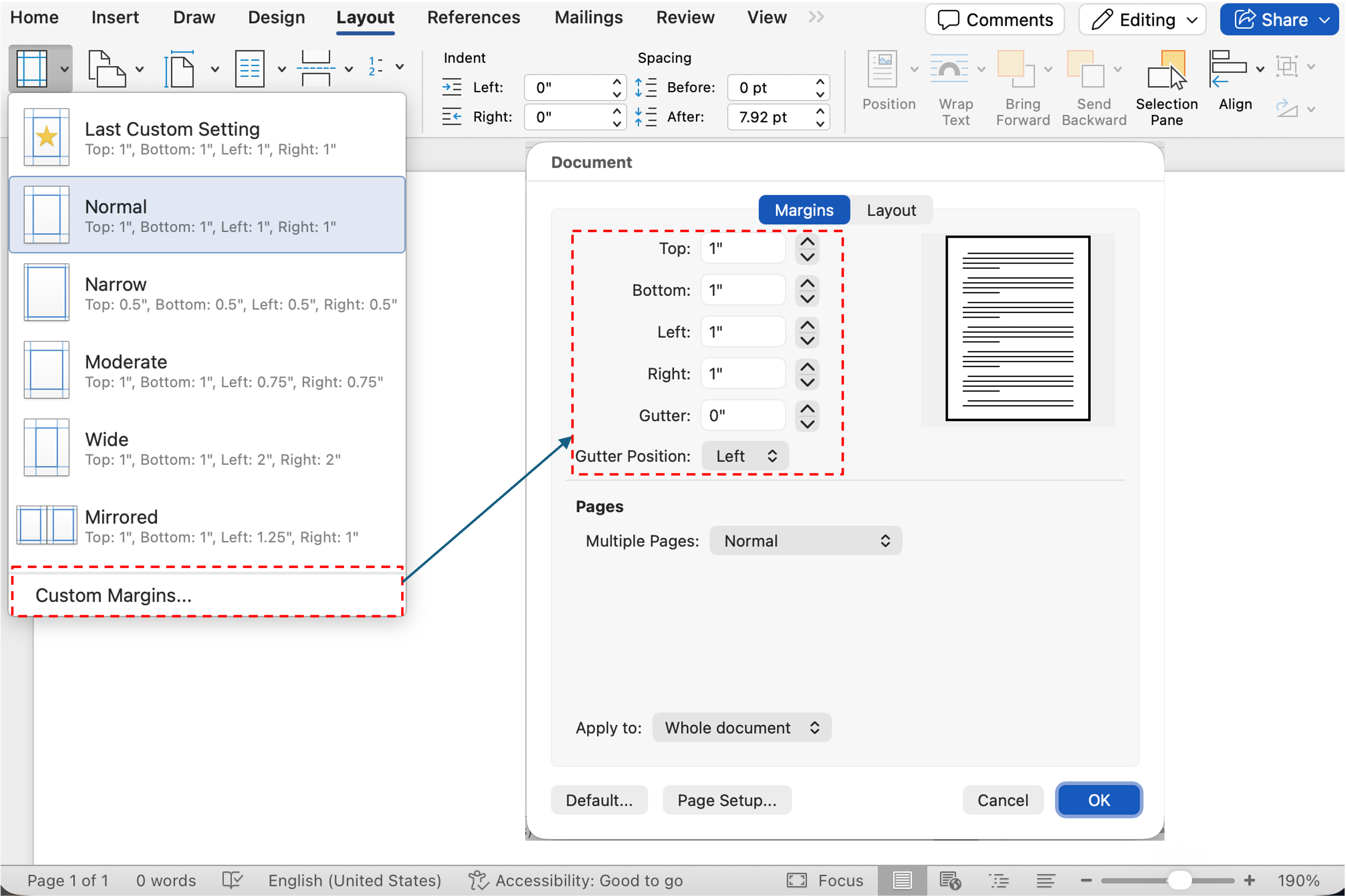}
        \caption{\textbf{Multi-step planning.}
        Configuring custom margins requires traversing
        \textit{Layout}, \textit{Margins}, and \textit{Custom Margins}
        before entering several parameter values. Although the necessary
        controls are available, the agent fails to construct the
        intermediate interaction sequence required to reach them.}
        \label{fig:failure-planning}
    \end{subfigure}
    \hfill
    \begin{subfigure}[t]{0.48\textwidth}
        \centering
        \includegraphics[width=\linewidth]{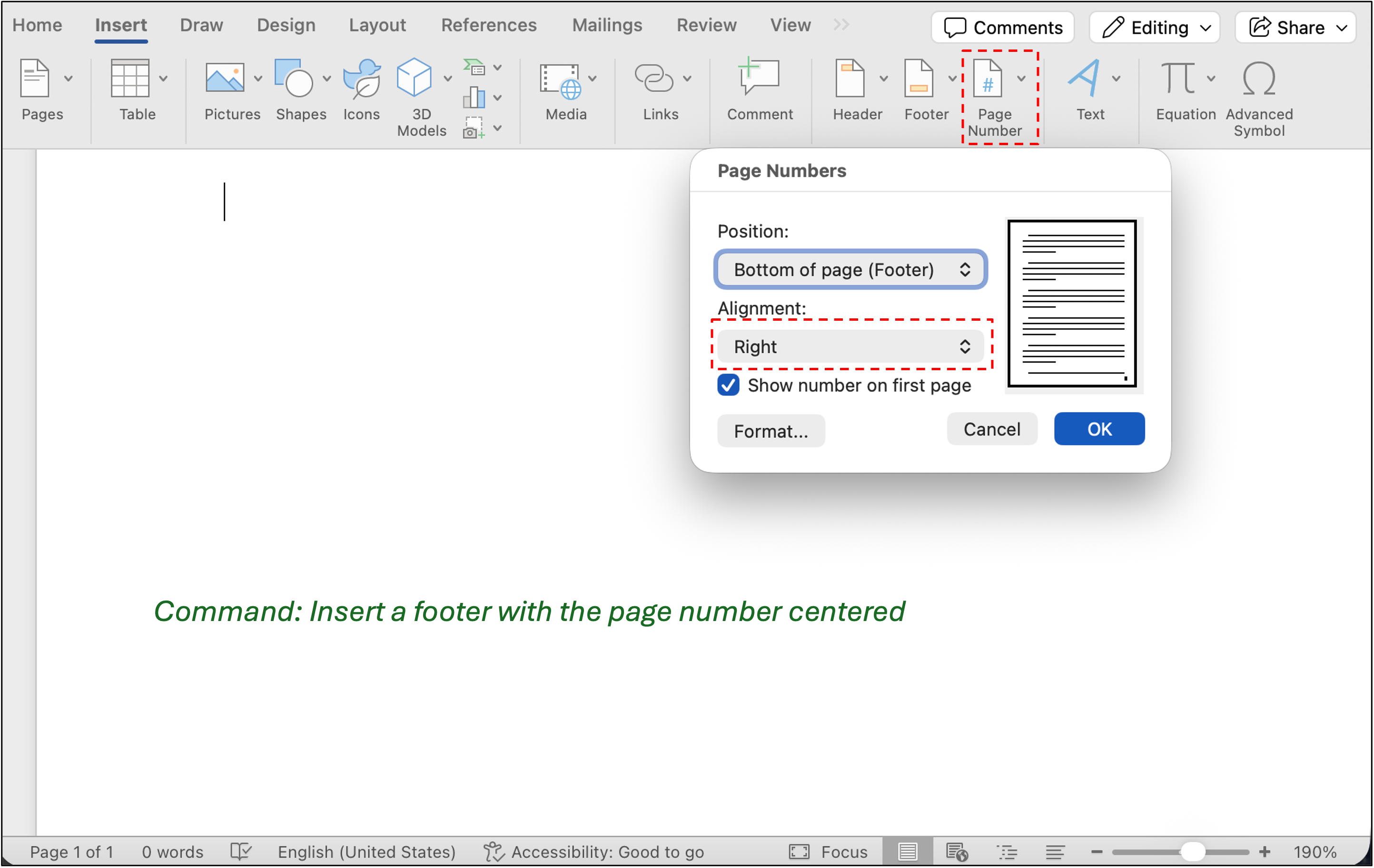}
        \caption{\textbf{Constraint binding.}
        For ``insert a footer with the page number centered,'' the agent
        reaches the page-number configuration and performs the principal
        insertion action, but applies \textit{Right} alignment instead
        of maintaining the requested \textit{centered} constraint.}
        \label{fig:failure-constraint}
    \end{subfigure}

    \caption{\textbf{Breakdowns during multi-step execution.}
    The left trace shows failure to derive the intermediate path needed
    to expose a valid configuration, while the right trace shows loss of
    a user-specified requirement after substantial task progress.}
    \label{fig:failure-examples-execution}
\end{figure}

\vspace{-65pt}
Figures~\ref{fig:failure-examples-grounding},
\ref{fig:failure-examples-reasoning}, and
\ref{fig:failure-examples-execution} visualize six breakdown patterns
identified in the RQ2 trace analysis, spanning interface grounding,
context maintenance, prior-knowledge reliance, default selection,
multi-step planning, and constraint binding.

\vspace{-45pt}

\vspace{-15pt}

%% file: main.bbl
\begin{thebibliography}{79}
\providecommand{\natexlab}[1]{#1}

\bibitem[{Anthropic(2025)}]{anthropic2025computeruse}
Anthropic. 2025.
\newblock \href
  {https://platform.claude.com/docs/en/agents-and-tools/tool-use/computer-use-tool}
  {Computer use tool documentation}.
\newblock Accessed: 2026-04-29.

\bibitem[{{Apple Inc.}(2026)}]{voiceover}
{Apple Inc.} 2026.
\newblock \href {https://support.apple.com/guide/voiceover/welcome/mac}
  {Voiceover user guide for mac}.
\newblock Accessed: 2026-04-23.

\bibitem[{Ashok(2018)}]{ashok2018non}
Vikas~Ganjigunte Ashok. 2018.
\newblock \emph{Non-Visual Web Browsing: From Accessibility with Screen Readers
  to Usability with Assistants}.
\newblock State University of New York at Stony Brook.

\bibitem[{Bai et~al.(2025)Bai, Cai, Chen, Chen, Chen, Cheng, Deng, Ding, Gao,
  Ge et~al.}]{bai2025qwen3}
Shuai Bai, Yuxuan Cai, Ruizhe Chen, Keqin Chen, Xionghui Chen, Zesen Cheng,
  Lianghao Deng, Wei Ding, Chang Gao, Chunjiang Ge, and 1 others. 2025.
\newblock Qwen3-vl technical report.
\newblock \emph{arXiv preprint arXiv:2511.21631}.

\bibitem[{Baldwin et~al.(2017)Baldwin, Hayes, Haimson, Mankoff, and
  Hudson}]{baldwin2017tangible}
Mark~S Baldwin, Gillian~R Hayes, Oliver~L Haimson, Jennifer Mankoff, and
  Scott~E Hudson. 2017.
\newblock The tangible desktop: a multimodal approach to nonvisual computing.
\newblock \emph{ACM Transactions on Accessible Computing (TACCESS)},
  10(3):1--28.

\bibitem[{Billah et~al.(2017)Billah, Ashok, Porter, and
  Ramakrishnan}]{billah2017ubiquitous}
Syed~Masum Billah, Vikas Ashok, Donald~E Porter, and IV~Ramakrishnan. 2017.
\newblock Ubiquitous accessibility for people with visual impairments: Are we
  there yet?
\newblock In \emph{Proceedings of the 2017 chi conference on human factors in
  computing systems}, pages 5862--5868.

\bibitem[{Bingham and Witkowsky(2021)}]{bingham2021deductive}
Andrea~J Bingham and Patricia Witkowsky. 2021.
\newblock Deductive and inductive approaches to qualitative data analysis.
\newblock \emph{Analyzing and interpreting qualitative data: After the
  interview}, 1:133--146.

\bibitem[{Bolger et~al.(2003)Bolger, Davis, and Rafaeli}]{bolger2003diary}
Niall Bolger, Angelina Davis, and Eshkol Rafaeli. 2003.
\newblock Diary methods: Capturing life as it is lived.
\newblock \emph{Annual review of psychology}, 54(1):579--616.

\bibitem[{Bonatti et~al.(2024)Bonatti, Zhao, Bonacci, Dupont, Abdali, Li, Lu,
  Wagle, Koishida, Bucker et~al.}]{bonatti2024windows}
Rogerio Bonatti, Dan Zhao, Francesco Bonacci, Dillon Dupont, Sara Abdali,
  Yinheng Li, Yadong Lu, Justin Wagle, Kazuhito Koishida, Arthur Bucker, and 1
  others. 2024.
\newblock Windows agent arena: Evaluating multi-modal os agents at scale.
\newblock \emph{arXiv preprint arXiv:2409.08264}.

\bibitem[{Braun and Clarke(2021)}]{braun2021thematic}
Virginia Braun and Victoria Clarke. 2021.
\newblock Thematic analysis: A practical guide.

\bibitem[{Caruana et~al.(2015)Caruana, Roman, Hern{\'a}ndez-S{\'a}nchez, and
  Solli}]{caruana2015longitudinal}
Edward~Joseph Caruana, Marius Roman, Jules Hern{\'a}ndez-S{\'a}nchez, and
  Piergiorgio Solli. 2015.
\newblock Longitudinal studies.
\newblock \emph{Journal of thoracic disease}, 7(11):E537.

\bibitem[{Chen et~al.(2026)Chen, Lu, Wang, Qiu, Chen, and
  Yang}]{chen2026struggle}
Nan Chen, Jing Lu, Zilong Wang, Luna~K Qiu, Siming Chen, and Yuqing Yang. 2026.
\newblock From struggle to success: Context-aware guidance for screen reader
  users in computer use.
\newblock In \emph{Proceedings of the 2026 CHI Conference on Human Factors in
  Computing Systems}, pages 1--19.

\bibitem[{Chezelles et~al.(2024)Chezelles, Le~Sellier, Shayegan, Jang, L{\`u},
  Yoran, Kong, Xu, Reddy, Cappart et~al.}]{chezelles2024browsergym}
De~Chezelles, Thibault Le~Sellier, Sahar~Omidi Shayegan, Lawrence~Keunho Jang,
  Xing~Han L{\`u}, Ori Yoran, Dehan Kong, Frank~F Xu, Siva Reddy, Quentin
  Cappart, and 1 others. 2024.
\newblock The browsergym ecosystem for web agent research.
\newblock \emph{arXiv preprint arXiv:2412.05467}.

\bibitem[{Christiano et~al.(2017)Christiano, Leike, Brown, Martic, Legg, and
  Amodei}]{christiano2017deep}
Paul~F Christiano, Jan Leike, Tom Brown, Miljan Martic, Shane Legg, and Dario
  Amodei. 2017.
\newblock Deep reinforcement learning from human preferences.
\newblock \emph{Advances in neural information processing systems}, 30.

\bibitem[{Coimbra and Dsilva(2026)}]{coimbra2026geminispark}
Adam Coimbra and Charmaine Dsilva. 2026.
\newblock Gemini spark now integrates with chrome.
\newblock
  \url{https://blog.google/innovation-and-ai/products/gemini-app/gemini-spark-updates-july-2026/}.
\newblock Google Blog, accessed 2026-08-30.

\bibitem[{Cui et~al.(2023)Cui, Karamcheti, Palleti, Shivakumar, Liang, and
  Sadigh}]{cui2023no}
Yuchen Cui, Siddharth Karamcheti, Raj Palleti, Nidhya Shivakumar, Percy Liang,
  and Dorsa Sadigh. 2023.
\newblock No, to the right: Online language corrections for robotic
  manipulation via shared autonomy.
\newblock In \emph{Proceedings of the 2023 ACM/IEEE International Conference on
  Human-Robot Interaction}, pages 93--101.

\bibitem[{Davydova et~al.(2025)Davydova, Jeffries, Barker, Flores, and
  Ryan}]{davydova2025osuniverse}
Mariya Davydova, Daniel Jeffries, Patrick Barker, Arturo~M{\'a}rquez Flores,
  and Sin{\'e}ad Ryan. 2025.
\newblock Osuniverse: Benchmark for multimodal gui-navigation ai agents.
\newblock \emph{arXiv preprint arXiv:2505.03570}.

\bibitem[{Deng et~al.(2023)Deng, Gu, Zheng, Chen, Stevens, Wang, Sun, and
  Su}]{deng2023mind2web}
Xiang Deng, Yu~Gu, Boyuan Zheng, Shijie Chen, Sam Stevens, Boshi Wang, Huan
  Sun, and Yu~Su. 2023.
\newblock Mind2web: Towards a generalist agent for the web.
\newblock \emph{Advances in Neural Information Processing Systems},
  36:28091--28114.

\bibitem[{Doush and Pontelli(2013)}]{doush2013non}
Iyad~Abu Doush and Enrico Pontelli. 2013.
\newblock Non-visual navigation of spreadsheets: Enhancing accessibility of
  microsoft excel™.
\newblock \emph{Universal access in the information society}, 12(2):143--159.

\bibitem[{Drouin et~al.(2024)Drouin, Gasse, Caccia, Laradji, Del~Verme, Marty,
  Boisvert, Thakkar, Cappart, Vazquez et~al.}]{drouin2024workarena}
Alexandre Drouin, Maxime Gasse, Massimo Caccia, Issam~H Laradji, Manuel
  Del~Verme, Tom Marty, L{\'e}o Boisvert, Megh Thakkar, Quentin Cappart, David
  Vazquez, and 1 others. 2024.
\newblock Workarena: How capable are web agents at solving common knowledge
  work tasks?
\newblock \emph{arXiv preprint arXiv:2403.07718}.

\bibitem[{Feng et~al.(2026)Feng, Chen, Wu, Zhou, and
  Bosselut}]{feng2026tracking}
Yiyang Feng, Zeming Chen, Haotian Wu, Jiawei Zhou, and Antoine Bosselut. 2026.
\newblock Tracking the limits of knowledge propagation: How llms fail at
  multi-step reasoning with conflicting knowledge.
\newblock In \emph{Proceedings of the 19th Conference of the European Chapter
  of the Association for Computational Linguistics (Volume 1: Long Papers)},
  pages 5813--5847.

\bibitem[{Feng et~al.(2025)Feng, Wang, Cui, Faltings, Lee, and
  Zhou}]{feng2025unraveling}
Yiyang Feng, Yichen Wang, Shaobo Cui, Boi Faltings, Mina Lee, and Jiawei Zhou.
  2025.
\newblock Unraveling misinformation propagation in llm reasoning.
\newblock In \emph{Findings of the Association for Computational Linguistics:
  EMNLP 2025}, pages 11683--11707.

\bibitem[{{Google}(2025)}]{google2025computeruse}
{Google}. 2025.
\newblock Gemini 2.5 computer use model.
\newblock
  \url{https://ai.google.dev/gemini-api/docs/models/gemini-2.5-computer-use-preview-10-2025}.
\newblock Accessed 2026-05-14.

\bibitem[{{Google DeepMind}(2025)}]{google2025projectastra}
{Google DeepMind}. 2025.
\newblock \href {https://deepmind.google/models/project-astra/} {Project
  astra}.
\newblock Accessed: 2026-04-29.

\bibitem[{Gubbi~Mohanbabu et~al.(2026)Gubbi~Mohanbabu, Natalie, Kim, Guo, and
  Pavel}]{gubbi2026a11y}
Ananya Gubbi~Mohanbabu, Rosiana Natalie, Brandon Kim, Anhong Guo, and Amy
  Pavel. 2026.
\newblock A11y-cua dataset: Characterizing the accessibility gap in computer
  use agents.
\newblock In \emph{Proceedings of the 2026 CHI Conference on Human Factors in
  Computing Systems}, pages 1--26.

\bibitem[{Harper and Chen(2012)}]{harper2012web}
Simon Harper and Alex~Q Chen. 2012.
\newblock Web accessibility guidelines: A lesson from the evolving web.
\newblock \emph{World Wide Web}, 15(1):61--88.

\bibitem[{He et~al.(2024)He, Yao, Ma, Yu, Dai, Zhang, Lan, and
  Yu}]{he2024webvoyager}
Hongliang He, Wenlin Yao, Kaixin Ma, Wenhao Yu, Yong Dai, Hongming Zhang,
  Zhenzhong Lan, and Dong Yu. 2024.
\newblock Webvoyager: Building an end-to-end web agent with large multimodal
  models.
\newblock In \emph{Proceedings of the 62nd Annual Meeting of the Association
  for Computational Linguistics (Volume 1: Long Papers)}, pages 6864--6890.

\bibitem[{Hsieh and Shannon(2005)}]{hsieh2005three}
Hsiu-Fang Hsieh and Sarah~E Shannon. 2005.
\newblock Three approaches to qualitative content analysis.
\newblock \emph{Qualitative health research}, 15(9):1277--1288.

\bibitem[{Islam et~al.(2023)Islam, Porter, and Billah}]{islam2023probabilistic}
Md~Touhidul Islam, Donald~E Porter, and Syed~Masum Billah. 2023.
\newblock A probabilistic model and metrics for estimating perceived
  accessibility of desktop applications in keystroke-based non-visual
  interactions.
\newblock In \emph{Proceedings of the 2023 CHI Conference on Human Factors in
  Computing Systems}, pages 1--20.

\bibitem[{Kodandaram et~al.(2023)Kodandaram, Sunkara, Jayarathna, and
  Ashok}]{kodandaram2023detecting}
Satwik~Ram Kodandaram, Mohan Sunkara, Sampath Jayarathna, and Vikas Ashok.
  2023.
\newblock Detecting deceptive dark-pattern web advertisements for blind
  screen-reader users.
\newblock \emph{Journal of Imaging}, 9(11):239.

\bibitem[{Kodandaram et~al.(2024)Kodandaram, Uckun, Bi, Ramakrishnan, and
  Ashok}]{kodandaram2024enabling}
Satwik~Ram Kodandaram, Utku Uckun, Xiaojun Bi, IV~Ramakrishnan, and Vikas
  Ashok. 2024.
\newblock Enabling uniform computer interaction experience for blind users
  through large language models.
\newblock In \emph{Proceedings of the 26th International ACM SIGACCESS
  Conference on Computers and Accessibility}, pages 1--14.

\bibitem[{Kodandaram et~al.(2026)Kodandaram, Zhou, Bi, Ramakrishnan, and
  Ashok}]{kodandaram2026finding}
Satwik~Ram Kodandaram, Jiawei Zhou, Xiaojun Bi, IV~Ramakrishnan, and Vikas
  Ashok. 2026.
\newblock Finding the signal in the noise: An exploratory study on assessing
  the effectiveness of ai and accessibility forums for blind users’ support
  needs.
\newblock In \emph{Proceedings of the 2026 CHI Conference on Human Factors in
  Computing Systems}, pages 1--20.

\bibitem[{Koh et~al.(2024)Koh, Lo, Jang, Duvvur, Lim, Huang, Neubig, Zhou,
  Salakhutdinov, and Fried}]{koh2024visualwebarena}
Jing~Yu Koh, Robert Lo, Lawrence Jang, Vikram Duvvur, Ming Lim, Po-Yu Huang,
  Graham Neubig, Shuyan Zhou, Russ Salakhutdinov, and Daniel Fried. 2024.
\newblock Visualwebarena: Evaluating multimodal agents on realistic visual web
  tasks.
\newblock In \emph{Proceedings of the 62nd Annual Meeting of the Association
  for Computational Linguistics (Volume 1: Long Papers)}, pages 881--905.

\bibitem[{Lan et~al.(2026)Lan, Sun, Walter, and Zhou}]{lan2026seeing}
Zixuan Lan, Luzhe Sun, Matthew~R Walter, and Jiawei Zhou. 2026.
\newblock Seeing without looking: Do vision-language benchmarks really test
  vision?
\newblock In \emph{Proceedings of the IEEE/CVF Conference on Computer Vision
  and Pattern Recognition}, pages 11260--11273.

\bibitem[{{LangChain}(2026)}]{langsmith2026}
{LangChain}. 2026.
\newblock Langsmith: Observability, evaluation, and deployment platform for ai
  agents.
\newblock \url{https://smith.langchain.com/}.
\newblock Agent engineering platform for debugging, testing, and monitoring
  LLM-based systems.

\bibitem[{Lazar et~al.(2017)Lazar, Feng, and Hochheiser}]{lazar2017research}
Jonathan Lazar, Jinjuan~Heidi Feng, and Harry Hochheiser. 2017.
\newblock \emph{Research methods in human-computer interaction}.
\newblock Morgan Kaufmann.

\bibitem[{Lee et~al.(2020)Lee, Ashok, and Ramakrishnan}]{lee2020repurposing}
Hae-Na Lee, Vikas Ashok, and IV~Ramakrishnan. 2020.
\newblock Repurposing visual input modalities for blind users: a case study of
  word processors.
\newblock In \emph{2020 IEEE International Conference on Systems, Man, and
  Cybernetics (SMC)}, pages 2714--2721. IEEE.

\bibitem[{Leporini et~al.(2012)Leporini, Buzzi, and
  Buzzi}]{leporini2012interacting}
Barbara Leporini, Maria~Claudia Buzzi, and Marina Buzzi. 2012.
\newblock Interacting with mobile devices via voiceover: usability and
  accessibility issues.
\newblock In \emph{Proceedings of the 24th Australian computer-human
  interaction conference}, pages 339--348.

\bibitem[{Levy et~al.(2024)Levy, Wiesel, Marreed, Oved, Yaeli, and
  Shlomov}]{levy2024st}
Ido Levy, Ben Wiesel, Sami Marreed, Alon Oved, Avi Yaeli, and Segev Shlomov.
  2024.
\newblock St-webagentbench: A benchmark for evaluating safety and
  trustworthiness in web agents.
\newblock \emph{arXiv preprint arXiv:2410.06703}.

\bibitem[{Li et~al.(2026)Li, Hu, Zheng, Zhou, and Chen}]{li2026orce}
Chen Li, Xiaoling Hu, Songzhu Zheng, Jiawei Zhou, and Chao Chen. 2026.
\newblock Orce: Order-aware alignment of verbalized confidence in large
  language models.
\newblock \emph{arXiv preprint arXiv:2605.12446}.

\bibitem[{Li et~al.(2025)Li, Xu, Tang, Livescu, McAllester, and
  Zhou}]{li2025okbench}
Yanhong Li, Tianyang Xu, Kenan Tang, Karen Livescu, David McAllester, and
  Jiawei Zhou. 2025.
\newblock Okbench: Democratizing llm evaluation with fully automated,
  on-demand, open knowledge benchmarking.
\newblock \emph{arXiv preprint arXiv:2511.08598}.

\bibitem[{Lightman et~al.(2024)Lightman, Kosaraju, Burda, Edwards, Baker, Lee,
  Leike, Schulman, Sutskever, and Cobbe}]{lightman2024let}
Hunter Lightman, Vineet Kosaraju, Yuri Burda, Harrison Edwards, Bowen Baker,
  Teddy Lee, Jan Leike, John Schulman, Ilya Sutskever, and Karl Cobbe. 2024.
\newblock Let's verify step by step.
\newblock In \emph{International Conference on Learning Representations},
  volume 2024, pages 39578--39601.

\bibitem[{Lin et~al.(2025{\natexlab{a}})Lin, Li, Gao, Yang, Wu, Bai, Lei, Wang,
  and Shou}]{lin2025showui}
Kevin~Qinghong Lin, Linjie Li, Difei Gao, Zhengyuan Yang, Shiwei Wu, Zechen
  Bai, Stan~Weixian Lei, Lijuan Wang, and Mike~Zheng Shou. 2025{\natexlab{a}}.
\newblock Showui: One vision-language-action model for gui visual agent.
\newblock In \emph{Proceedings of the Computer Vision and Pattern Recognition
  Conference}, pages 19498--19508.

\bibitem[{Lin et~al.(2025{\natexlab{b}})Lin, Zhou, and Yu}]{lin2025llm}
Samuel Lin, Jiawei Zhou, and Minlan Yu. 2025{\natexlab{b}}.
\newblock An llm-based agentic framework for accessible networkcontrol.
\newblock \emph{ACM SIGMETRICS Performance Evaluation Review}, 53(2):15--20.

\bibitem[{Miao et~al.(2016)Miao, Pham, Friebe, and Weber}]{miao2016contrasting}
Mei Miao, Hoai~Anh Pham, Jens Friebe, and Gerhard Weber. 2016.
\newblock Contrasting usability evaluation methods with blind users.
\newblock \emph{Universal access in the Information Society}, 15(1):63--76.

\bibitem[{{Microsoft}(2026{\natexlab{a}})}]{microsoft2026copilot}
{Microsoft}. 2026{\natexlab{a}}.
\newblock \href {https://learn.microsoft.com/en-us/copilot/} {Microsoft copilot
  overview}.
\newblock Accessed: 2026-04-29.

\bibitem[{{Microsoft}(2026{\natexlab{b}})}]{microsoft_uiautomation_win32}
{Microsoft}. 2026{\natexlab{b}}.
\newblock Microsoft ui automation: Accessibility framework for windows desktop
  applications.
\newblock
  \url{https://learn.microsoft.com/en-us/windows/win32/winauto/entry-uiauto-win32}.
\newblock Provides programmatic access to UI elements for accessibility and
  automation; accessed 2026-05-06.

\bibitem[{Morales et~al.(2013)Morales, Arteaga, and
  Kurniawan}]{morales2013design}
Lourdes Morales, Sonia~M Arteaga, and Sri Kurniawan. 2013.
\newblock Design guidelines of a tool to help blind authors independently
  format their word documents.
\newblock In \emph{CHI'13 Extended Abstracts on Human Factors in Computing
  Systems}, pages 31--36.

\bibitem[{Naeem et~al.(2023)Naeem, Ozuem, Howell, and Ranfagni}]{naeem2023step}
Muhammad Naeem, Wilson Ozuem, Kerry Howell, and Silvia Ranfagni. 2023.
\newblock A step-by-step process of thematic analysis to develop a conceptual
  model in qualitative research.
\newblock \emph{International journal of qualitative methods},
  22:16094069231205789.

\bibitem[{Noy(2008)}]{noy2008sampling}
Chaim Noy. 2008.
\newblock Sampling knowledge: The hermeneutics of snowball sampling in
  qualitative research.
\newblock \emph{International Journal of social research methodology},
  11(4):327--344.

\bibitem[{{NV Access}(2026)}]{nvaccess}
{NV Access}. 2026.
\newblock \href {https://www.nvaccess.org/} {Nv access}.
\newblock Accessed: 2026-04-23.

\bibitem[{OpenAI(2025)}]{openai2025operator}
OpenAI. 2025.
\newblock Introducing operator.
\newblock \url{https://openai.com/index/introducing-operator/}.
\newblock Accessed: 2026-04-29.

\bibitem[{Ouyang et~al.(2022)Ouyang, Wu, Jiang, Almeida, Wainwright, Mishkin,
  Zhang, Agarwal, Slama, Ray et~al.}]{ouyang2022training}
Long Ouyang, Jeffrey Wu, Xu~Jiang, Diogo Almeida, Carroll Wainwright, Pamela
  Mishkin, Chong Zhang, Sandhini Agarwal, Katarina Slama, Alex Ray, and 1
  others. 2022.
\newblock Training language models to follow instructions with human feedback.
\newblock \emph{Advances in neural information processing systems},
  35:27730--27744.

\bibitem[{Peng et~al.(2025)Peng, Li, Bigham, and Pavel}]{peng2025morae}
Yi-Hao Peng, Dingzeyu Li, Jeffrey~P Bigham, and Amy Pavel. 2025.
\newblock Morae: Proactively pausing ui agents for user choices.
\newblock In \emph{Proceedings of the 38th Annual ACM Symposium on User
  Interface Software and Technology}, pages 1--14.

\bibitem[{Perera et~al.(2026)Perera, Ananthanarayan, Goncu, and
  Marriott}]{perera2026m}
Minoli Perera, Swamy Ananthanarayan, Cagatay Goncu, and Kim Marriott. 2026.
\newblock I'm always a little skeptical of it: Verification practices of blind
  users when working with generative ai in spreadsheets.
\newblock In \emph{Proceedings of the 2026 CHI Conference on Human Factors in
  Computing Systems}, pages 1--21.

\bibitem[{Qin et~al.(2025)Qin, Ye, Fang, Wang, Liang, Tian, Zhang, Li, Li,
  Huang et~al.}]{qin2025ui}
Yujia Qin, Yining Ye, Junjie Fang, Haoming Wang, Shihao Liang, Shizuo Tian,
  Junda Zhang, Jiahao Li, Yunxin Li, Shijue Huang, and 1 others. 2025.
\newblock Ui-tars: Pioneering automated gui interaction with native agents.
\newblock \emph{arXiv preprint arXiv:2501.12326}.

\bibitem[{Rawles et~al.()Rawles, Clinckemaillie, Chang, Waltz, Lau, Fair, Li,
  Bishop, Li, Campbell-Ajala et~al.}]{rawles2405androidworld}
Christopher Rawles, Sarah Clinckemaillie, Yifan Chang, Jonathan Waltz,
  Gabrielle Lau, Marybeth Fair, Alice Li, William Bishop, Wei Li, Folawiyo
  Campbell-Ajala, and 1 others.
\newblock Androidworld: A dynamic benchmarking environment for autonomous
  agents, 2024.
\newblock \emph{URL https://arxiv. org/abs/2405.14573}.

\bibitem[{Rawles et~al.(2023)Rawles, Li, Rodriguez, Riva, and
  Lillicrap}]{rawles2023androidinthewild}
Christopher Rawles, Alice Li, Daniel Rodriguez, Oriana Riva, and Timothy
  Lillicrap. 2023.
\newblock Androidinthewild: A large-scale dataset for android device control.
\newblock \emph{Advances in Neural Information Processing Systems},
  36:59708--59728.

\bibitem[{Ross et~al.(2011)Ross, Gordon, and Bagnell}]{ross2011reduction}
St{\'e}phane Ross, Geoffrey Gordon, and Drew Bagnell. 2011.
\newblock A reduction of imitation learning and structured prediction to
  no-regret online learning.
\newblock In \emph{Proceedings of the fourteenth international conference on
  artificial intelligence and statistics}, pages 627--635. JMLR Workshop and
  Conference Proceedings.

\bibitem[{Shayegani et~al.(2025)Shayegani, Hines, Dong, Abu-Ghazaleh, Lutz,
  Whitehead, Balachandran, Nushi, and Vineet}]{shayegani2025just}
Erfan Shayegani, Keegan Hines, Yue Dong, Nael Abu-Ghazaleh, Roman Lutz, Spencer
  Whitehead, Vidhisha Balachandran, Besmira Nushi, and Vibhav Vineet. 2025.
\newblock Just do it!? computer-use agents exhibit blind goal-directedness.
\newblock \emph{arXiv preprint arXiv:2510.01670}.

\bibitem[{Singh et~al.(2025)Singh, Fry, Perelman, Tart, Ganesh, El-Kishky,
  McLaughlin, Low, Ostrow, Ananthram et~al.}]{singh2025openai}
Aaditya Singh, Adam Fry, Adam Perelman, Adam Tart, Adi Ganesh, Ahmed El-Kishky,
  Aidan McLaughlin, Aiden Low, AJ~Ostrow, Akhila Ananthram, and 1 others. 2025.
\newblock Openai gpt-5 system card.
\newblock \emph{arXiv preprint arXiv:2601.03267}.

\bibitem[{Song et~al.(2025)Song, Su, Qu, Zhou, and Cheng}]{song2025prmbench}
Mingyang Song, Zhaochen Su, Xiaoye Qu, Jiawei Zhou, and Yu~Cheng. 2025.
\newblock Prmbench: A fine-grained and challenging benchmark for process-level
  reward models.
\newblock In \emph{Proceedings of the 63rd Annual Meeting of the Association
  for Computational Linguistics (Volume 1: Long Papers)}, pages 25299--25346.

\bibitem[{Sunkara et~al.(2023)Sunkara, Kalari, Jayarathna, and
  Ashok}]{sunkara2023assessing}
Mohan Sunkara, Sandeep Kalari, Sampath Jayarathna, and Vikas Ashok. 2023.
\newblock Assessing the accessibility of web archives.
\newblock In \emph{2023 ACM/IEEE Joint Conference on Digital Libraries (JCDL)},
  pages 253--255. IEEE.

\bibitem[{Uckun et~al.(2022)Uckun, Tumkur~Suresh, Ferdous, Bi, Ramakrishnan,
  and Ashok}]{uckun2022taming}
Utku Uckun, Rohan Tumkur~Suresh, Md~Javedul Ferdous, Xiaojun Bi,
  IV~Ramakrishnan, and Vikas Ashok. 2022.
\newblock Taming user-interface heterogeneity with uniform overlays for blind
  users.
\newblock In \emph{Proceedings of the 30th ACM conference on user modeling,
  adaptation and personalization}, pages 212--222.

\bibitem[{{Vispero}(2026)}]{jaws}
{Vispero}. 2026.
\newblock \href {https://vispero.com/jaws-screen-reader-software/} {Jaws screen
  reader software}.
\newblock Accessed: 2026-04-23.

\bibitem[{Wentz et~al.(2013)Wentz, Hochheiser, and Lazar}]{wentz2013survey}
Brian Wentz, Harry Hochheiser, and Jonathan Lazar. 2013.
\newblock A survey of blind users on the usability of email applications.
\newblock \emph{Universal access in the information society}, 12(3):327--336.

\bibitem[{Wentz and Lazar(2011)}]{wentz2011usability}
Brian Wentz and Jonathan Lazar. 2011.
\newblock Usability evaluation of email applications by blind users.
\newblock \emph{Journal of Usability Studies}, 6(2):75--89.

\bibitem[{{World Wide Web Consortium}(2023)}]{waiaria12}
{World Wide Web Consortium}. 2023.
\newblock {Accessible Rich Internet Applications (WAI-ARIA) 1.2}.
\newblock \url{https://www.w3.org/TR/wai-aria-1.2/}.
\newblock W3C Recommendation, 6 June 2023.

\bibitem[{{World Wide Web Consortium}(2024)}]{wcag22}
{World Wide Web Consortium}. 2024.
\newblock {Web Content Accessibility Guidelines (WCAG) 2.2}.
\newblock \url{https://www.w3.org/TR/WCAG22/}.
\newblock W3C Recommendation, updated 12 December 2024.

\bibitem[{Xie et~al.(2024)Xie, Zhang, Chen, Li, Zhao, Cao, Hua, Cheng, Shin,
  Lei et~al.}]{xie2024osworld}
Tianbao Xie, Danyang Zhang, Jixuan Chen, Xiaochuan Li, Siheng Zhao, Ruisheng
  Cao, Toh~J Hua, Zhoujun Cheng, Dongchan Shin, Fangyu Lei, and 1 others. 2024.
\newblock Osworld: Benchmarking multimodal agents for open-ended tasks in real
  computer environments.
\newblock \emph{Advances in Neural Information Processing Systems},
  37:52040--52094.

\bibitem[{Xu et~al.(2026)Xu, Song, Li, Tang, Jain, Bao, Wang, Zhou, Guo, Cao
  et~al.}]{xu2026theagentcompany}
Frank~Fangzheng Xu, Yufan Song, Boxuan Li, Yuxuan Tang, Kritanjali Jain,
  Mengxue Bao, Zora Wang, Xuhui Zhou, Zhitong Guo, Murong Cao, and 1 others.
  2026.
\newblock Theagentcompany: benchmarking llm agents on consequential real world
  tasks.
\newblock \emph{Advances in Neural Information Processing Systems}, 38.

\bibitem[{Xue et~al.(2025)Xue, Qi, Shi, Song, Gou, Song, Sun, and
  Su}]{xue2025illusion}
Tianci Xue, Weijian Qi, Tianneng Shi, Chan~Hee Song, Boyu Gou, Dawn Song, Huan
  Sun, and Yu~Su. 2025.
\newblock An illusion of progress? assessing the current state of web agents.
\newblock \emph{arXiv preprint arXiv:2504.01382}.

\bibitem[{Yang et~al.(2026)Yang, Shirkavand, Jin, Zhou, Gao, and
  Huang}]{yang2026capability}
Haoyan Yang, Reza Shirkavand, Yukai Jin, Jiawei Zhou, Shangqian Gao, and Heng
  Huang. 2026.
\newblock Capability self-assessment: Teaching llms to know their limits.
\newblock \emph{arXiv preprint arXiv:2606.00251}.

\bibitem[{Yao et~al.(2022)Yao, Chen, Yang, and Narasimhan}]{yao2022webshop}
Shunyu Yao, Howard Chen, John Yang, and Karthik Narasimhan. 2022.
\newblock Webshop: Towards scalable real-world web interaction with grounded
  language agents.
\newblock \emph{Advances in Neural Information Processing Systems},
  35:20744--20757.

\bibitem[{Zhang et~al.(2025)Zhang, Li, He, Zhang, Qiao, Qin, Ma, Kang, Lin,
  Rajmohan et~al.}]{zhang2025ufo}
Chaoyun Zhang, Liqun Li, Shilin He, Xu~Zhang, Bo~Qiao, Si~Qin, Minghua Ma,
  Yu~Kang, Qingwei Lin, Saravan Rajmohan, and 1 others. 2025.
\newblock Ufo: A ui-focused agent for windows os interaction.
\newblock In \emph{Proceedings of the 2025 Conference of the Nations of the
  Americas Chapter of the Association for Computational Linguistics: Human
  Language Technologies (Volume 1: Long Papers)}, pages 597--622.

\bibitem[{Zhou(2026)}]{zhou2026position}
Jiawei Zhou. 2026.
\newblock Position: Scores without context? rethinking the role of evaluation
  in the era of llms.
\newblock In \emph{Proceedings of the Fifth Workshop on Generation, Evaluation
  and Metrics (GEM)}, pages 1048--1054.

\bibitem[{Zhou et~al.(2022)Zhou, Eisner, Newman, Platanios, and
  Thomson}]{zhou2022online}
Jiawei Zhou, Jason Eisner, Michael Newman, Emmanouil~Antonios Platanios, and
  Sam Thomson. 2022.
\newblock Online semantic parsing for latency reduction in task-oriented
  dialogue.
\newblock In \emph{Proceedings of the 60th Annual Meeting of the Association
  for Computational Linguistics (Volume 1: Long Papers)}, pages 1554--1576.

\bibitem[{Zhou et~al.(2023)Zhou, Xu, Zhu, Zhou, Lo, Sridhar, Cheng, Ou, Bisk,
  Fried et~al.}]{zhou2023webarena}
Shuyan Zhou, Frank~F Xu, Hao Zhu, Xuhui Zhou, Robert Lo, Abishek Sridhar,
  Xianyi Cheng, Tianyue Ou, Yonatan Bisk, Daniel Fried, and 1 others. 2023.
\newblock Webarena: A realistic web environment for building autonomous agents.
\newblock \emph{arXiv preprint arXiv:2307.13854}.

\bibitem[{Ziegler et~al.(2019)Ziegler, Stiennon, Wu, Brown, Radford, Amodei,
  Christiano, and Irving}]{ziegler2019fine}
Daniel~M Ziegler, Nisan Stiennon, Jeffrey Wu, Tom~B Brown, Alec Radford, Dario
  Amodei, Paul Christiano, and Geoffrey Irving. 2019.
\newblock Fine-tuning language models from human preferences.
\newblock \emph{arXiv preprint arXiv:1909.08593}.

\end{thebibliography}
